%Paper: astro-ph/9510089
%From: griest@astrophys.ucsd.edu (Kim Griest)
%Date: Tue, 17 Oct 95 17:51:18 PDT
%Date (revised): Tue, 17 Oct 95 18:05:14 PDT

% write-up of  Varenna Lectures July 1995
% written:  July 30 1995
% last updated: 12 Sep 1995
%%% FIRST COMES THE ENTIRE PHYZZX MACRO PACKAGE %%%%
%%% SKIP TO "START OF PAPER" to find start of paper
%%%%%%%%%%%%%%%%%%%%%%%%%%%%%%%%%%%%%%%%%%%%%%%%%%%%%%%%%%%%%%%%%%%%%%%%
% % % % % % % % % % % % % % % % % % % % % % % % % % % % % % % % % % % %
%%%   This is PHYZZX macro package.   % % % % % % % % % % % % % % % % %
%% % % % % % % % % % % % % % % % % % % % % % % % % % % % % % % % % % % %
%%%  This version of PHYZZX should be used with Version 1.0 of TEX  % %
%% % % % % % % % % % % % % % % % % % % % % % % % % % % % % % % % % % % %
%%%   Do not "\input phyzzx" unless you preload or "\input" PLAIN.  % %
%% % % % % % % % % % % % % % % % % % % % % % % % % % % % % % % % % % % %
%%%   To preload both PLAIN and PHYZZX, begin your file with    % % % %
%%%  a line "%macropackage=phyzzx" instead of "\input phyzzx".  % % % %
%% % % % % % % % % % % % % % % % % % % % % % % % % % % % % % % % % % % %
%%%%%%%%%%%%%%%%%%%%%%%%%%%%%%%%%%%%%%%%%%%%%%%%%%%%%%%%%%%%%%%%%%%%%%%%
%%%%%%%  Created by Vadim Kaplunovsky in June 1984.   %%%%%%%%%%%%%%%%%%
% % % % % % % % % % % % % % % % % % % % % % % % % % % % % % % % % % % %
%%%%%%%%%%%%  Latest update/debug: September 24, 1984.    %%%%%%%%%%%%%%
%%%%%%%%%%%%%%%%%%%%%%%%%%%%%%%%%%%%%%%%%%%%%%%%%%%%%%%%%%%%%%%%%%%%%%%%
%
\catcode`@=11 % This allows us to modify PLAIN macros.
%
%%%%%%%%%%%%%%%%%%%%%%%%%%%%%%%%%%%%%%%%%%%%%%%%%%%%%%%%%%%%%%%%%%%%%%%%
%
%   I begin with fonts.
%

\font\fourteenrm=cmr10 scaled\magstep2
\font\twelverm=cmr10 scaled\magstep1
\font\ninerm=cmr9            \font\sixrm=cmr6

\font\fourteenbf=cmbx10 scaled\magstep2
\font\twelvebf=cmbx10 scaled\magstep1
\font\ninebf=cmbx9            \font\sixbf=cmbx6
\font\seventeeni=cmmi10 scaled\magstep3     \skewchar\seventeeni='177
\font\fourteeni=cmmi10 scaled\magstep2      \skewchar\fourteeni='177
\font\twelvei=cmmi10 scaled\magstep1        \skewchar\twelvei='177
\font\ninei=cmmi9                           \skewchar\ninei='177
\font\sixi=cmmi6                            \skewchar\sixi='177
\font\seventeensy=cmsy10 scaled\magstep3    \skewchar\seventeensy='60
\font\fourteensy=cmsy10 scaled\magstep2     \skewchar\fourteensy='60
\font\twelvesy=cmsy10 scaled\magstep1       \skewchar\twelvesy='60
\font\ninesy=cmsy9                          \skewchar\ninesy='60
\font\sixsy=cmsy6                           \skewchar\sixsy='60

\font\fourteenex=cmex10 scaled\magstep2
\font\twelveex=cmex10 scaled\magstep1

\font\fourteensl=cmsl10 scaled\magstep2
\font\twelvesl=cmsl10 scaled\magstep1
\font\ninesl=cmsl9

\font\fourteenit=cmti10 scaled\magstep2
\font\twelveit=cmti10 scaled\magstep1
\font\twelvett=cmtt10 scaled\magstep1
\font\twelvecp=cmcsc10 scaled\magstep1
\font\tencp=cmcsc10
\newfam\cpfam
%
      % quick fix for a missing font
%
\newcount\f@ntkey            \f@ntkey=0
\def\samef@nt{\relax \ifcase\f@ntkey \rm \or\oldstyle \or\or
         \or\it \or\sl \or\bf \or\tt \or\caps \fi }
\def\fourteenpoint{\relax
    \textfont0=\fourteenrm          \scriptfont0=\tenrm
    \scriptscriptfont0=\sevenrm
     \def\rm{\fam0 \fourteenrm \f@ntkey=0 }\relax
    \textfont1=\fourteeni           \scriptfont1=\teni
    \scriptscriptfont1=\seveni
     \def\oldstyle{\fam1 \fourteeni\f@ntkey=1 }\relax
    \textfont2=\fourteensy          \scriptfont2=\tensy
    \scriptscriptfont2=\sevensy
    \textfont3=\fourteenex     \scriptfont3=\fourteenex
    \scriptscriptfont3=\fourteenex
    \def\it{\fam\itfam \fourteenit\f@ntkey=4 }\textfont\itfam=\fourteenit
    \def\sl{\fam\slfam \fourteensl\f@ntkey=5 }\textfont\slfam=\fourteensl
    \scriptfont\slfam=\tensl
    \def\bf{\fam\bffam \fourteenbf\f@ntkey=6 }\textfont\bffam=\fourteenbf
    \scriptfont\bffam=\tenbf     \scriptscriptfont\bffam=\sevenbf
    \def\tt{\fam\ttfam \twelvett \f@ntkey=7 }\textfont\ttfam=\twelvett
    \h@big=11.9\p@{} \h@Big=16.1\p@{} \h@bigg=20.3\p@{} \h@Bigg=24.5\p@{}
    \def\caps{\fam\cpfam \twelvecp \f@ntkey=8 }\textfont\cpfam=\twelvecp
    \setbox\strutbox=\hbox{\vrule height 12pt depth 5pt width\z@}
    \samef@nt}
\def\twelvepoint{\relax
    \textfont0=\twelverm          \scriptfont0=\ninerm
    \scriptscriptfont0=\sixrm
     \def\rm{\fam0 \twelverm \f@ntkey=0 }\relax
    \textfont1=\twelvei           \scriptfont1=\ninei
    \scriptscriptfont1=\sixi
     \def\oldstyle{\fam1 \twelvei\f@ntkey=1 }\relax
    \textfont2=\twelvesy          \scriptfont2=\ninesy
    \scriptscriptfont2=\sixsy
    \textfont3=\twelveex          \scriptfont3=\twelveex
    \scriptscriptfont3=\twelveex
    \def\it{\fam\itfam \twelveit \f@ntkey=4 }\textfont\itfam=\twelveit
    \def\sl{\fam\slfam \twelvesl \f@ntkey=5 }\textfont\slfam=\twelvesl
    \scriptfont\slfam=\ninesl
    \def\bf{\fam\bffam \twelvebf \f@ntkey=6 }\textfont\bffam=\twelvebf
    \scriptfont\bffam=\ninebf     \scriptscriptfont\bffam=\sixbf
    \def\tt{\fam\ttfam \twelvett \f@ntkey=7 }\textfont\ttfam=\twelvett
    \h@big=10.2\p@{}
    \h@Big=13.8\p@{}
    \h@bigg=17.4\p@{}
    \h@Bigg=21.0\p@{}
    \def\caps{\fam\cpfam \twelvecp \f@ntkey=8 }\textfont\cpfam=\twelvecp
    \setbox\strutbox=\hbox{\vrule height 10pt depth 4pt width\z@}
    \samef@nt}
\def\tenpoint{\relax
    \textfont0=\tenrm          \scriptfont0=\sevenrm
    \scriptscriptfont0=\fiverm
    \def\rm{\fam0 \tenrm \f@ntkey=0 }\relax
    \textfont1=\teni           \scriptfont1=\seveni
    \scriptscriptfont1=\fivei
    \def\oldstyle{\fam1 \teni \f@ntkey=1 }\relax
    \textfont2=\tensy          \scriptfont2=\sevensy
    \scriptscriptfont2=\fivesy
    \textfont3=\tenex          \scriptfont3=\tenex
    \scriptscriptfont3=\tenex
    \def\it{\fam\itfam \tenit \f@ntkey=4 }\textfont\itfam=\tenit
    \def\sl{\fam\slfam \tensl \f@ntkey=5 }\textfont\slfam=\tensl
    \def\bf{\fam\bffam \tenbf \f@ntkey=6 }\textfont\bffam=\tenbf
    \scriptfont\bffam=\sevenbf     \scriptscriptfont\bffam=\fivebf
    \def\tt{\fam\ttfam \tentt \f@ntkey=7 }\textfont\ttfam=\tentt
    \def\caps{\fam\cpfam \tencp \f@ntkey=8 }\textfont\cpfam=\tencp
    \setbox\strutbox=\hbox{\vrule height 8.5pt depth 3.5pt width\z@}
    \samef@nt}
%
%%%%%%%%%%%%%%%%%%%%%%%%%%%%%%%%%%%%%%%%%%%%%%%%%%%%%%%%%%%%%%%%%%%%%%%%
%
%   Next redifine \big \Big \bigg and \Bigg to work with all fonts
%
%%%%%%%%%%%%%%%%%%%%%%%%%%%%%%%%%%%%%%%%%%%%%%%%%%%%%%%%%%%%%%%%%%%%%%%%
%
\newdimen\h@big  \h@big=8.5\p@
\newdimen\h@Big  \h@Big=11.5\p@
\newdimen\h@bigg  \h@bigg=14.5\p@
\newdimen\h@Bigg  \h@Bigg=17.5\p@
\def\big#1{{\hbox{$\left#1\vbox to\h@big{}\right.\n@space$}}}
\def\Big#1{{\hbox{$\left#1\vbox to\h@Big{}\right.\n@space$}}}
\def\bigg#1{{\hbox{$\left#1\vbox to\h@bigg{}\right.\n@space$}}}
\def\Bigg#1{{\hbox{$\left#1\vbox to\h@Bigg{}\right.\n@space$}}}
%
%%%%%%%%%%%%%%%%%%%%%%%%%%%%%%%%%%%%%%%%%%%%%%%%%%%%%%%%%%%%%%%%%%%%%%%%
%
%   Next, I define basic spacing parameters.
%
\normalbaselineskip = 20pt plus 0.2pt minus 0.1pt
\normallineskip = 1.5pt plus 0.1pt minus 0.1pt
\normallineskiplimit = 1.5pt
\newskip\normaldisplayskip
\normaldisplayskip = 20pt plus 5pt minus 10pt
\newskip\normaldispshortskip
\normaldispshortskip = 6pt plus 5pt
\newskip\normalparskip
\normalparskip = 6pt plus 2pt minus 1pt
\newskip\skipregister
\skipregister = 5pt plus 2pt minus 1.5pt
\newif\ifsingl@    \newif\ifdoubl@
\newif\iftwelv@    \twelv@true
\def\singlespace{\singl@true\doubl@false\spaces@t}
\def\doublespace{\singl@false\doubl@true\spaces@t}
\def\normalspace{\singl@false\doubl@false\spaces@t}
\def\Tenpoint{\tenpoint\twelv@false\spaces@t}
\def\Twelvepoint{\twelvepoint\twelv@true\spaces@t}
\def\spaces@t{\relax%
 \iftwelv@ \ifsingl@\subspaces@t3:4;\else\subspaces@t1:1;\fi%
 \else \ifsingl@\subspaces@t3:5;\else\subspaces@t4:5;\fi \fi%
 \ifdoubl@ \multiply\baselineskip by 5%
 \divide\baselineskip by 4 \fi \unskip}
\def\subspaces@t#1:#2;{
      \baselineskip = \normalbaselineskip
      \multiply\baselineskip by #1 \divide\baselineskip by #2
      \lineskip = \normallineskip
      \multiply\lineskip by #1 \divide\lineskip by #2
      \lineskiplimit = \normallineskiplimit
      \multiply\lineskiplimit by #1 \divide\lineskiplimit by #2
      \parskip = \normalparskip
      \multiply\parskip by #1 \divide\parskip by #2
      \abovedisplayskip = \normaldisplayskip
      \multiply\abovedisplayskip by #1 \divide\abovedisplayskip by #2
      \belowdisplayskip = \abovedisplayskip
      \abovedisplayshortskip = \normaldispshortskip
      \multiply\abovedisplayshortskip by #1
        \divide\abovedisplayshortskip by #2
      \belowdisplayshortskip = \abovedisplayshortskip
      \advance\belowdisplayshortskip by \belowdisplayskip
      \divide\belowdisplayshortskip by 2
      \smallskipamount = \skipregister
      \multiply\smallskipamount by #1 \divide\smallskipamount by #2
      \medskipamount = \smallskipamount \multiply\medskipamount by 2
      \bigskipamount = \smallskipamount \multiply\bigskipamount by 4 }
\def\normalbaselines{ \baselineskip=\normalbaselineskip
   \lineskip=\normallineskip \lineskiplimit=\normallineskip
   \iftwelv@\else \multiply\baselineskip by 4 \divide\baselineskip by 5
     \multiply\lineskiplimit by 4 \divide\lineskiplimit by 5
     \multiply\lineskip by 4 \divide\lineskip by 5 \fi }
\Twelvepoint  % That's the default
\interlinepenalty=50
\interfootnotelinepenalty=5000
\predisplaypenalty=9000
\postdisplaypenalty=500
\hfuzz=1pt
\vfuzz=0.2pt
%
%%%%%%%%%%%%%%%%%%%%%%%%%%%%%%%%%%%%%%%%%%%%%%%%%%%%%%%%%%%%%%%%%%%%%%%%
%
%   Next, I define output routines, footnotes & related stuff.
%
\def\pagecontents{
   \ifvoid\topins\else\unvbox\topins\vskip\skip\topins\fi
   \dimen@ = \dp255 \unvbox255
   \ifvoid\footins\else\vskip\skip\footins\footrule\unvbox\footins\fi
   \ifr@ggedbottom \kern-\dimen@ \vfil \fi }
\def\makeheadline{\vbox to 0pt{ \skip@=\topskip
      \advance\skip@ by -12pt \advance\skip@ by -2\normalbaselineskip
      \vskip\skip@ \line{\vbox to 12pt{}\the\headline} \vss
      }\nointerlineskip}
\def\makefootline{\baselineskip = 1.5\normalbaselineskip
                 \line{\the\footline}}
\newif\iffrontpage
\newif\ifletterstyle
\newif\ifp@genum
\def\nopagenumbers{\p@genumfalse}
\def\pagenumbers{\p@genumtrue}
\pagenumbers
\newtoks\paperheadline
\newtoks\letterheadline
\newtoks\letterfrontheadline
\newtoks\lettermainheadline
\newtoks\paperfootline
\newtoks\letterfootline
\newtoks\date
\footline={\ifletterstyle\the\letterfootline\else\the\paperfootline\fi}
\paperfootline={\hss\iffrontpage\else\ifp@genum\tenrm\folio\hss\fi\fi}
\letterfootline={\hfil}
\headline={\ifletterstyle\the\letterheadline\else\the\paperheadline\fi}
\paperheadline={\hfil}
\letterheadline{\iffrontpage\the\letterfrontheadline
     \else\the\lettermainheadline\fi}
\lettermainheadline={\rm\ifp@genum page \ \folio\fi\hfil\the\date}
\def\monthname{\relax\ifcase\month 0/\or January\or February\or
   March\or April\or May\or June\or July\or August\or September\or
   October\or November\or December\else\number\month/\fi}
\date={\monthname\ \number\day, \number\year}
\countdef\pagenumber=1  \pagenumber=1
\def\advancepageno{\global\advance\pageno by 1
   \ifnum\pagenumber<0 \global\advance\pagenumber by -1
    \else\global\advance\pagenumber by 1 \fi \global\frontpagefalse }
\def\folio{\ifnum\pagenumber<0 \romannumeral-\pagenumber
           \else \number\pagenumber \fi }
\def\footrule{\dimen@=\prevdepth\nointerlineskip
   \vbox to 0pt{\vskip -0.25\baselineskip \hrule width 0.35\hsize \vss}
   \prevdepth=\dimen@ }
\newtoks\foottokens
\foottokens={\Tenpoint\singlespace}
\newdimen\footindent
\footindent=24pt
\def\vfootnote#1{\insert\footins\bgroup  \the\foottokens
   \interlinepenalty=\interfootnotelinepenalty \floatingpenalty=20000
   \splittopskip=\ht\strutbox \boxmaxdepth=\dp\strutbox
   \leftskip=\footindent \rightskip=\z@skip
   \parindent=0.5\footindent \parfillskip=0pt plus 1fil
   \spaceskip=\z@skip \xspaceskip=\z@skip
   \Textindent{$ #1 $}\footstrut\futurelet\next\fo@t}
\def\Textindent#1{\noindent\llap{#1\enspace}\ignorespaces}
\def\footnote#1{\attach{#1}\vfootnote{#1}}

\def\foot{\attach\footsymbolgen\vfootnote{\footsymbol}}
\let\footsymbol=\star
\newcount\lastf@@t           \lastf@@t=-1
\newcount\footsymbolcount    \footsymbolcount=0
\newif\ifPhysRev
\def\footsymbolgen{\relax \ifPhysRev \iffrontpage \NPsymbolgen\else
      \PRsymbolgen\fi \else \NPsymbolgen\fi
   \global\lastf@@t=\pageno \footsymbol }
\def\NPsymbolgen{\ifnum\footsymbolcount<0 \global\footsymbolcount=0\fi
   {\iffrontpage \else \advance\lastf@@t by 1 \fi
    \ifnum\lastf@@t<\pageno \global\footsymbolcount=0
     \else \global\advance\footsymbolcount by 1 \fi }
   \ifcase\footsymbolcount \fd@f\star\or \fd@f\dagger\or \fd@f\ast\or
    \fd@f\ddagger\or \fd@f\natural\or \fd@f\diamond\or \fd@f\bullet\or
    \fd@f\nabla\else \fd@f\dagger\global\footsymbolcount=0 \fi }
\def\fd@f#1{\xdef\footsymbol{#1}}
\def\PRsymbolgen{\ifnum\footsymbolcount>0 \global\footsymbolcount=0\fi
      \global\advance\footsymbolcount by -1
      \xdef\footsymbol{\sharp\number-\footsymbolcount} }
\def\space@ver#1{\let\@sf=\empty \ifmmode #1\else \ifhmode
   \edef\@sf{\spacefactor=\the\spacefactor}\unskip${}#1$\relax\fi\fi}
\def\attach#1{\space@ver{\strut^{\mkern 2mu #1} }\@sf\ }
%
%%%%%%%%%%%%%%%%%%%%%%%%%%%%%%%%%%%%%%%%%%%%%%%%%%%%%%%%%%%%%%%%%%%%%%%%
%
%   Here come chapter, section, subsection & appendix macros.
%
\newcount\chapternumber      \chapternumber=0
\newcount\sectionnumber      \sectionnumber=0
\newcount\equanumber         \equanumber=0
\let\chapterlabel=0
\newtoks\chapterstyle        \chapterstyle={\Number}
\newskip\chapterskip         \chapterskip=\bigskipamount
\newskip\sectionskip         \sectionskip=\medskipamount
\newskip\headskip            \headskip=8pt plus 3pt minus 3pt
\newdimen\chapterminspace    \chapterminspace=15pc
\newdimen\sectionminspace    \sectionminspace=10pc
\newdimen\referenceminspace  \referenceminspace=25pc
\def\chapterreset{\global\advance\chapternumber by 1
   \ifnum\the\equanumber<0 \else\global\equanumber=0\fi
   \sectionnumber=0 \makel@bel}
\def\makel@bel{\xdef\chapterlabel{%
\the\chapterstyle{\the\chapternumber}.}}
\def\sectionlabel{\number\sectionnumber \quad }
\def\alphabetic#1{\count255='140 \advance\count255 by #1\char\count255}
\def\Alphabetic#1{\count255='100 \advance\count255 by #1\char\count255}
\def\Roman#1{\uppercase\expandafter{\romannumeral #1}}
\def\roman#1{\romannumeral #1}
\def\Number#1{\number #1}
\def\unnumberedchapters{\let\makel@bel=\relax \let\chapterlabel=\relax
\let\sectionlabel=\relax \equanumber=-1 }
\def\titlestyle#1{\par\begingroup \interlinepenalty=9999
     \leftskip=0.02\hsize plus 0.23\hsize minus 0.02\hsize
     \rightskip=\leftskip \parfillskip=0pt
     \hyphenpenalty=9000 \exhyphenpenalty=9000
     \tolerance=9999 \pretolerance=9000
     \spaceskip=0.333em \xspaceskip=0.5em
     \iftwelv@\fourteenpoint\else\twelvepoint\fi
   \noindent #1\par\endgroup }
\def\spacecheck#1{\dimen@=\pagegoal\advance\dimen@ by -\pagetotal
   \ifdim\dimen@<#1 \ifdim\dimen@>0pt \vfil\break \fi\fi}
\def\chapter#1{\par \penalty-300 \vskip\chapterskip
   \spacecheck\chapterminspace
   \chapterreset \titlestyle{\chapterlabel \ #1}
   \nobreak\vskip\headskip \penalty 30000
   \wlog{\string\chapter\ \chapterlabel} }

\def\section#1{\par \ifnum\the\lastpenalty=30000\else
   \penalty-200\vskip\sectionskip \spacecheck\sectionminspace\fi
   \wlog{\string\section\ \chapterlabel \the\sectionnumber}
   \global\advance\sectionnumber by 1  \noindent
   {\caps\enspace\chapterlabel \sectionlabel #1}\par
   \nobreak\vskip\headskip \penalty 30000 }
\def\subsection#1{\par
   \ifnum\the\lastpenalty=30000\else \penalty-100\smallskip \fi
   \noindent\undertext{#1}\enspace \vadjust{\penalty5000}}

\def\undertext#1{\vtop{\hbox{#1}\kern 1pt \hrule}}
\def\APPENDIX#1#2{\par\penalty-300\vskip\chapterskip
   \spacecheck\chapterminspace \chapterreset \xdef\chapterlabel{#1}
   \titlestyle{APPENDIX #2} \nobreak\vskip\headskip \penalty 30000
   \wlog{\string\Appendix\ \chapterlabel} }
\def\Appendix#1{\APPENDIX{#1}{#1}}
\def\appendix{\APPENDIX{A}{}}
%
%%%%%%%%%%%%%%%%%%%%%%%%%%%%%%%%%%%%%%%%%%%%%%%%%%%%%%%%%%%%%%%%%%%%%%%%
%
%   Here come macros for equation numbering.
%
\def\eqname#1{\relax \ifnum\the\equanumber<0
     \xdef#1{{\rm(\number-\equanumber)}}\global\advance\equanumber by -1
    \else \global\advance\equanumber by 1
      \xdef#1{{\rm(\chapterlabel \number\equanumber)}} \fi}
\def\eqinsert#1{\noalign{\dimen@=\prevdepth \nointerlineskip
   \setbox0=\hbox to\displaywidth{\hfil #1}
   \vbox to 0pt{\vss\hbox{$\!\box0\!$}\kern-0.5\baselineskip}
   \prevdepth=\dimen@}}
%

%

%

%
%%%%%%%%%%%%%%%%%%%%%%%%%%%%%%%%%%%%%%%%%%%%%%%%%%%%%%%%%%%%%%%%%%%%%%%%
%   Here come items and lists
%
\def\GENITEM#1;#2{\par \hangafter=0 \hangindent=#1
    \Textindent{$ #2 $}\ignorespaces}
\outer\def\newitem#1=#2;{\gdef#1{\GENITEM #2;}}
\newdimen\itemsize                \itemsize=30pt
\newitem\item=1\itemsize;
\newitem\sitem=1.75\itemsize;     
\newitem\ssitem=2.5\itemsize;     
\outer\def\newlist#1=#2&#3&#4;{\toks0={#2}\toks1={#3}%
   \count255=\escapechar \escapechar=-1
   \alloc@0\list\countdef\insc@unt\listcount     \listcount=0
   \edef#1{\par
      \countdef\listcount=\the\allocationnumber
      \advance\listcount by 1
      \hangafter=0 \hangindent=#4
      \Textindent{\the\toks0{\listcount}\the\toks1}}
   \expandafter\expandafter\expandafter
    \edef\c@t#1{begin}{\par
      \countdef\listcount=\the\allocationnumber \listcount=1
      \hangafter=0 \hangindent=#4
      \Textindent{\the\toks0{\listcount}\the\toks1}}
   \expandafter\expandafter\expandafter
    \edef\c@t#1{con}{\par \hangafter=0 \hangindent=#4 \noindent}
   \escapechar=\count255}
\def\c@t#1#2{\csname\string#1#2\endcsname}
\newlist\point=\Number&.&1.0\itemsize;
\newlist\subpoint=(\alphabetic&)&1.75\itemsize;
\newlist\subsubpoint=(\roman&)&2.5\itemsize;
\newlist\cpoint=\Roman&.&1.0\itemsize;
%

%
%%%%%%%%%%%%%%%%%%%%%%%%%%%%%%%%%%%%%%%%%%%%%%%%%%%%%%%%%%%%%%%%%%%%%%%%
%
%   Here come macros for references, figures & tables.
%
\newcount\referencecount     \referencecount=0
\newif\ifreferenceopen       \newwrite\referencewrite
\newtoks\rw@toks
\def\NPrefmark#1{\attach{\scriptscriptstyle [ #1 ] }}
\let\PRrefmark=\attach
\def\CErefmark#1{\attach{\scriptstyle  #1 ) }}
\def\refmark#1{\relax\ifPhysRev\PRrefmark{#1}\else\NPrefmark{#1}\fi}
\def\crefmark#1{\relax\CErefmark{#1}}
\def\refend{\refmark{\number\referencecount}}
\newcount\lastrefsbegincount \lastrefsbegincount=0
\def\refsend{\refmark{\count255=\referencecount
   \advance\count255 by-\lastrefsbegincount
   \ifcase\count255 \number\referencecount
   \or \number\lastrefsbegincount,\number\referencecount
   \else \number\lastrefsbegincount-\number\referencecount \fi}}
\def\crefsend{\crefmark{\count255=\referencecount
   \advance\count255 by-\lastrefsbegincount
   \ifcase\count255 \number\referencecount
   \or \number\lastrefsbegincount,\number\referencecount
   \else \number\lastrefsbegincount-\number\referencecount \fi}}
\def\refch@ck{\chardef\rw@write=\referencewrite
   \ifreferenceopen \else \referenceopentrue
   \immediate\openout\referencewrite=referenc.texauxil \fi}
%
% In \obeyendofline, we say `\let^^M=\relax
{\catcode`\^^M=\active % these lines must end with %
  \gdef\obeyendofline{\catcode`\^^M\active \let^^M\ }}%
%
% In \ignoreendofline, we say `\let^^M=\relax
{\catcode`\^^M=\active % these lines must end with %
  \gdef\ignoreendofline{\catcode`\^^M=5}}
{\obeyendofline\gdef\rw@start#1{\def\t@st{#1} \ifx\t@st\blankend%
\endgroup \@sf \relax \else \ifx\t@st\bl@nkend \endgroup \@sf \relax%
\else \rw@begin#1
\backtotext
\fi \fi } }
{\obeyendofline\gdef\rw@begin#1
{\def\n@xt{#1}\rw@toks={#1}\relax%
\rw@next}}
\def\blankend{}
{\obeylines\gdef\bl@nkend{
}}
\newif\iffirstrefline  \firstreflinetrue
\def\rwr@teswitch{\ifx\n@xt\blankend \let\n@xt=\rw@begin %
 \else\iffirstrefline \global\firstreflinefalse%
\immediate\write\rw@write{\noexpand\obeyendofline \the\rw@toks}%
\let\n@xt=\rw@begin%
      \else\ifx\n@xt\rw@@d \def\n@xt{\immediate\write\rw@write{%
        \noexpand\ignoreendofline}\endgroup \@sf}%
             \else \immediate\write\rw@write{\the\rw@toks}%
             \let\n@xt=\rw@begin\fi\fi \fi}
\def\rw@next{\rwr@teswitch\n@xt}
\def\rw@@d{\backtotext} \let\rw@end=\relax
\let\backtotext=\relax

\newdimen\refindent     \refindent=30pt
\def\refitem#1{\par \hangafter=0 \hangindent=\refindent \Textindent{#1}}
\def\REFNUM#1{\space@ver{}\refch@ck \firstreflinetrue%
 \global\advance\referencecount by 1 \xdef#1{\the\referencecount}}
\def\refnum#1{\space@ver{}\refch@ck \firstreflinetrue%
 \global\advance\referencecount by 1 \xdef#1{\the\referencecount}\refend}

\def\REF#1{\REFNUM#1%
 \immediate\write\referencewrite{%
 \noexpand\refitem{#1.}}%
\begingroup\obeyendofline\rw@start}
\def\ref{\refnum\?%
 \immediate\write\referencewrite{\noexpand\refitem{\?.}}%
\begingroup\obeyendofline\rw@start}
\def\Ref#1{\refnum#1%
 \immediate\write\referencewrite{\noexpand\refitem{#1.}}%
\begingroup\obeyendofline\rw@start}
\def\REFS#1{\REFNUM#1\global\lastrefsbegincount=\referencecount
\immediate\write\referencewrite{\noexpand\refitem{#1.}}%
\begingroup\obeyendofline\rw@start}
\newcount\figurecount     \figurecount=0
\newif\iffigureopen       \newwrite\figurewrite
\def\figch@ck{\chardef\rw@write=\figurewrite \iffigureopen\else
   \immediate\openout\figurewrite=figures.texauxil
   \figureopentrue\fi}
\def\FIGNUM#1{\space@ver{}\figch@ck \firstreflinetrue%
 \global\advance\figurecount by 1 \xdef#1{\the\figurecount}}
\def\FIG#1{\FIGNUM#1
   \immediate\write\figurewrite{\noexpand\refitem{#1.}}%
   \begingroup\obeyendofline\rw@start}
\def\fig{\FIGNUM\? fig.~\?%
\immediate\write\figurewrite{\noexpand\refitem{\?.}}%
\begingroup\obeyendofline\rw@start}
\def\figure{\FIGNUM\? figure~\?
   \immediate\write\figurewrite{\noexpand\refitem{\?.}}%
   \begingroup\obeyendofline\rw@start}
\def\Fig{\FIGNUM\? Fig.~\?%
\immediate\write\figurewrite{\noexpand\refitem{\?.}}%
\begingroup\obeyendofline\rw@start}
\def\Figure{\FIGNUM\? Figure~\?%
\immediate\write\figurewrite{\noexpand\refitem{\?.}}%
\begingroup\obeyendofline\rw@start}
\newcount\tablecount     \tablecount=0
\newif\iftableopen       \newwrite\tablewrite
\def\tabch@ck{\chardef\rw@write=\tablewrite \iftableopen\else
   \immediate\openout\tablewrite=tables.texauxil
   \tableopentrue\fi}
\def\TABNUM#1{\space@ver{}\tabch@ck \firstreflinetrue%
 \global\advance\tablecount by 1 \xdef#1{\the\tablecount}}
\def\TABLE#1{\TABNUM#1
   \immediate\write\tablewrite{\noexpand\refitem{#1.}}%
   \begingroup\obeyendofline\rw@start}
\def\Table{\TABNUM\? Table~\?%
\immediate\write\tablewrite{\noexpand\refitem{\?.}}%
\begingroup\obeyendofline\rw@start}
%

%
%%%%%%%%%%%%%%%%%%%%%%%%%%%%%%%%%%%%%%%%%%%%%%%%%%%%%%%%%%%%%%%%%%%%%%%%
%
%   Here come macros for memos & letters.
%
\def\masterreset{\global\pagenumber=1 \global\chapternumber=0
   \ifnum\the\equanumber<0\else \global\equanumber=0\fi
   \global\sectionnumber=0
   \global\referencecount=0 \global\figurecount=0 \global\tablecount=0 }
\def\FRONTPAGE{\ifvoid255\else\vfill\penalty-2000\fi
      \masterreset\global\frontpagetrue
      \global\lastf@@t=0 \global\footsymbolcount=0}

\def\paperstyle{\letterstylefalse\normalspace\papersize}
\def\letterstyle{\letterstyletrue\singlespace\lettersize}
\def\papersize{\hsize=35pc\vsize=48pc\hoffset=1pc\voffset=6pc
               \skip\footins=\bigskipamount}
\def\lettersize{\hsize=6.5in\vsize=8.5in\hoffset=0in\voffset=1in
   \skip\footins=\smallskipamount \multiply\skip\footins by 3 }
\paperstyle   %  This is the default
%
% % % % % % % % % % % % % % % % % % % % % % % % % % % % % % % % % % % %
%
\def\MEMO{\letterstyle\FRONTPAGE \letterfrontheadline={\hfil}
    \line{\quad\fourteenrm FNAL MEMORANDUM\hfil\twelverm\the\date\quad}
    \medskip \memod@f}

\def\memit@m#1{\smallskip \hangafter=0 \hangindent=1in
      \Textindent{\caps #1}}
\def\memod@f{\xdef\to{\memit@m{To:}}\xdef\from{\memit@m{From:}}%
     \xdef\topic{\memit@m{Topic:}}\xdef\subject{\memit@m{Subject:}}%
     \xdef\rule{\bigskip\hrule height 1pt\bigskip}}
\memod@f
\newskip\lettertopfil
\lettertopfil = 0pt plus 1.5in minus 0pt
\newskip\letterbottomfil
\letterbottomfil = 0pt plus 2.3in minus 0pt
\newskip\spskip \setbox0\hbox{\ } \spskip=-1\wd0
\def\addressee#1{\medskip\rightline{\the\date\hskip 30pt} \bigskip
   \vskip\lettertopfil
   \ialign to\hsize{\strut ##\hfil\tabskip 0pt plus \hsize \cr #1\crcr}
   \medskip\noindent\hskip\spskip}
\newskip\signatureskip       \signatureskip=40pt
\def\signed#1{\par \penalty 9000 \bigskip \dt@pfalse
  \everycr={\noalign{\ifdt@p\vskip\signatureskip\global\dt@pfalse\fi}}
  \setbox0=\vbox{\singlespace \halign{\tabskip 0pt \strut ##\hfil\cr
   \noalign{\global\dt@ptrue}#1\crcr}}
  \line{\hskip 0.5\hsize minus 0.5\hsize \box0\hfil} \medskip }

\def\endletter{\ifnum\pagenumber=1 \vskip\letterbottomfil\supereject
\else \vfil\supereject \fi}
\newbox\letterb@x
\def\lettertext{\par\unvcopy\letterb@x\par}
\def\multiletter{\setbox\letterb@x=\vbox\bgroup
      \everypar{\vrule height 1\baselineskip depth 0pt width 0pt }
      \singlespace \topskip=\baselineskip }
\def\letterend{\par\egroup}
%
%%%%%%%%%%%%%%%%%%%%%%%%%%%%%%%%%%%%%%%%%%%%%%%%%%%%%%%%%%%%%%%%%%%%%%%
%
%   Here come macros for title pages.
%
\newskip\frontpageskip
\newtoks\pubtype
\newtoks\Pubnum
\newtoks\pubnum
\newif\ifp@bblock  \p@bblocktrue
\def\PH@SR@V{\doubl@true \baselineskip=24.1pt plus 0.2pt minus 0.1pt
             \parskip= 3pt plus 2pt minus 1pt }
\def\PHYSREV{\paperstyle\PhysRevtrue\PH@SR@V}
\def\titlepage{\FRONTPAGE\paperstyle\ifPhysRev\PH@SR@V\fi
   \ifp@bblock\p@bblock\fi}
\def\nopubblock{\p@bblockfalse}

\frontpageskip=1\medskipamount plus .5fil
\pubtype={\tensl Preliminary Version}
%\Pubnum={$\caps FERMILAB - Pub - \the\pubnum $}
%\Pubnum={$\rm FERMILAB-Pub-\the\pubnum $}
\pubnum={0000}
\def\p@bblock{\begingroup \tabskip=\hsize minus \hsize
   \baselineskip=1.5\ht\strutbox \topspace-2\baselineskip
   \halign to\hsize{\strut ##\hfil\tabskip=0pt\crcr
%   \the\Pubnum\cr \the\date\cr \the\pubtype\cr}\endgroup}
   \the\Pubnum\cr \the\date\cr}\endgroup}
%   \the\date\cr}\endgroup}

%
\def\title#1{\vskip\frontpageskip \titlestyle{#1} \vskip\headskip }
\def\author#1{\vskip\frontpageskip\titlestyle{\twelvecp #1}\nobreak}

\def\address#1{\par\kern 5pt\titlestyle{\twelvepoint\it #1}}
\def\andaddress{\par\kern 5pt \centerline{\sl and} \address}

\def\abstract{\vskip\frontpageskip\centerline{\fourteenrm ABSTRACT}
              \vskip\headskip }

%
%
%%%%%%%%%%%%%%%%%%%%%%%%%%%%%%%%%%%%%%%%%%%%%%%%%%%%%%%%%%%%%%%%%%%%%%%%
%   Miscellaneous macros
%

\def\\{\relax\ifmmode\backslash\else$\backslash$\fi}
\def\globaleqnumbers{\relax\ifnum\the\equanumber<0%
\else\global\equanumber=-1\fi}

\def\journal#1&#2(#3){\unskip, \sl #1~\bf #2 \rm (19#3) }

\def\topspace{\hrule height 0pt depth 0pt \vskip}

\def\VEV#1{\left\langle #1\right\rangle}

\let\int=\intop         
\def\prop{\mathrel{{\mathchoice{\pr@p\scriptstyle}{\pr@p\scriptstyle}{
                \pr@p\scriptscriptstyle}{\pr@p\scriptscriptstyle} }}}
\def\pr@p#1{\setbox0=\hbox{$\cal #1 \char'103$}
   \hbox{$\cal #1 \char'117$\kern-.4\wd0\box0}}
\def\lsim{\mathrel{\mathpalette\@versim<}}
\def\gsim{\mathrel{\mathpalette\@versim>}}
\def\@versim#1#2{\lower0.2ex\vbox{\baselineskip\z@skip\lineskip\z@skip
  \lineskiplimit\z@\ialign{$\m@th#1\hfil##\hfil$\crcr#2\crcr\sim\crcr}}}
\def\leftrightarrowfill{$\m@th \mathord- \mkern-6mu
	\cleaders\hbox{$\mkern-2mu \mathord- \mkern-2mu$}\hfil
	\mkern-6mu \mathord\leftrightarrow$}
\def\lrover#1{\vbox{\ialign{##\crcr
	\leftrightarrowfill\crcr\noalign{\kern-1pt\nointerlineskip}
	$\hfil\displaystyle{#1}\hfil$\crcr}}}
%
% % % % % % % % % % % % % % % % % % % % % % % % % % % % % % % % % % % %
%
%   Finally, some bug fixings.
%
\let\sec@nt=\sec
\def\sec{\relax\ifmmode\let\n@xt=\sec@nt\else\let\n@xt\section\fi\n@xt}
\def\obsolete#1{\message{Macro \string #1 is obsolete.}}
\def\firstsec#1{\obsolete\firstsec \section{#1}}
\def\firstsubsec#1{\obsolete\firstsubsec \subsection{#1}}
\def\thispage#1{\obsolete\thispage \global\pagenumber=#1\frontpagefalse}
\def\thischapter#1{\obsolete\thischapter \global\chapternumber=#1}
\def\nextequation#1{\obsolete\nextequation \global\equanumber=#1
   \ifnum\the\equanumber>0 \global\advance\equanumber by 1 \fi}
\def\BOXITEM{\afterassigment\B@XITEM\setbox0=}
\def\B@XITEM{\par\hangindent\wd0 \noindent\box0 }
%

%%%%%%%%%%%%%%%%%%%%%%%%%%%%%%%%%%%%%%%%%%%%%%%%%%%%%%%%%%%%%%%%%%%%%%%%
%   That's about it
%
\catcode`@=12 % at signs are no longer letters
\message{ by V.K.}
\everyjob{\input myphyx }
\def\etal{{\it et al.}}
\vsize=8.75truein
\hsize=6.0truein
\voffset=-.3truein
%\nopagenumbers
\baselineskip=15pt
\overfullrule=0pt
\unnumberedchapters
%
%%%%%%%% definitions %%%%%%%%%%%%
\def\inunit{ {\rm m}^{-2} {\rm yr}^{-1}}
\def\Omh{\Omega_\chi h^2}
\def\la{<}
\def\ga{>}
\def\kmsec{km sec$^{-1}$}
\def\ten#1{\times 10^{#1}}

\def\msun{M_\odot}
\def\lsun{L_\odot}

\def\pac{Paczy{\'n}ski}
\def\etal{{\it et al.}}

\def\that{{\widehat t}}
   %%% These are milder than thesis

\def\rrr#1{[{#1}]}

%%%%% end of definitions %%%%%%%%
%%%%%%%% references %%%%%%%%%%%%%%%%
\REF\ashman{Ashman, K.M., 1992. PASP, 104, 1109.}
\REF\binney{Binney, J. \& Tremaine, S. 1987,
	Galactic Dynamics (Princeton University Press, Princeton)}
\REF\blitz{Blitz, L., Fich, M., and Stark, A. A., ApJ Supp. 49, 183 (1982)}.
\REF\merrifield{Merrifield, M.R. AJ, 103, 1552 (1992).}
\REF\zaritsky{ Zaritsky, D. \etal, 1989, ApJ, 345, 759.}
\REF\kochanek{C.S.Kochanek, e-print, astro-ph 9505068 (1995).}
\REF\tyson{J.A.Tyson \& P.Fischer, ApJ Lett. 446, L55 (1995).}
\REF\briel{U.~G.~Briel, J.~P.~Henry, \& H.~Bohringer, Astr. Astrophys.
	 259, L31 (1992).}
\REF\white{White, S.D.M., \etal\ Nature, 366, 429 (1993).}
\REF\Dekel{Dekel, A, 1994, ARA\&A, 32, 371.}
\REF\yahiletal{A. Yahil, T. Walker, and M. Rowan-Robinson,
     Astrophys. J. Lett. {\bf 301}, L1 (1986).}
\REF\davisnusser{M. Davis and A. Nusser, astro-ph/9501025.}
\REF\bbn{K. A. Olive, G.~Steigman, D.~N.~Schramm, T.~P.~Walker,
      and H.~Kang, {\rm Astrophys. J.} {\bf 376}, 51 (1991).}
\REF\smithkawmal{M. S. Smith, L. H. Kawano, and R. A. Malaney,
     Astrophys. J. Suppl. {\bf 85}, 219 (1993).}
\REF\walkeretal{T. P. Walker \etal, Astrophys. J. {\bf 376}, 51 (1991).}
\REF\depaolis{For example, F.DePaolis, \etal, A\&A 295, 567 (1995);
	F.DePaolis, \etal, Phys. Rev. Lett. 74, 14 (1995).}
\REF\mond{Begeman, K.G., Broeils, A.H., \& Sanders, R.H. 1991. \sl
     M.N.R.A.S. \bf 249, \rm 532.}
\REF\axion{for example, Turner, M.S. 1990, \sl Physics Reports \bf 197, \rm 67;
     Raffelt, G.G. 1990, \sl Physics Reports \bf 198, \rm 1.}
\REF\vanbibber{K.~Van Bibber, \etal, Int. J. Mod. Phys. D3, 33 (1994).}
\REF\hagmann{C.~Hagmann, \etal, Phys. Rev. D42, 1297 (1990).}
\REF\kolb{Kolb, E.W. \& Turner, M.S. 1990. \sl The Early Universe, \rm
     (Addison-Wesley, Redwood City, California).}
\REF\Jungman{Jungman, G. Kamionkowski, M., \& Griest, K., 1995, to appear
        in Physics Reports.}
\REF\direct{Proceedings of the 5th Workshop on Low Temperature Devices,
	Berkeley 1993, in J. Low Temp. Phys. 93, 185 (1993);
	J.R.Primack, B.Sadoulet, \& D.Seckel, Ann.Rev.Nucl.Part.Sci., B38,
	751 (1988); P.F.Smith \& J.D.Lewin, Phys. Rep. 187, 203 (1990).}
\REF\griestsadoulet{Griest, K. \& Sadoulet, B., in \sl Dark Matter in the
     Universe, \rm eds. Galeotti, P., \& Schramm, D.N. (Kluwer, Netherlands,
     1989).}
\REF\CDMS{P.D.Barnes \etal, J. Low Temp. Phys. 93, 79 (1993); T.Shutt \etal,
	Phys. Rev. Lett. 65, 1305 (1992); {\it ibid.} 3531 (1992).}
\REF\angela{Data acquired by the
	UCB/UCSB/LBL experiment at Oroville kindly provided by A.~Da Silva,
	unpublished.}
\REF\caldwell{D.~O.~ Caldwell \etal, {\rm Phys.~Rev.~Lett.}
       {\bf 61} (1988) 510.}
\REF\otherdirect{for example, N.Spooner \& P.F.Smith, Phys. Lett. B314,
	430 (1993); A.Bottino \etal, Phys. Lett. B293, 460 (1992);
	K.Fushimi, \etal, Phys. Rev. C47, 425 (1993); G.J.Davies \etal,
	Phys. Lett. B320, 395 (1994).}
\REF\indirect{M.Mori \etal\ (Kamionkande), Phys. Rev. D48, 5505 (1993);
	J.M.LoSecco \etal\ (IMB), Phys. Lett. B188, 388 (1987);
	E.Diehl, (MACRO) Ph.D thesis, University of Michegan (1994).}
\REF\newindirect{R.J.Wilkes, in proceedings of 22nd SLAC Summer Institute
	on Particle Physics, Stanford 1994; D.M.Lowder \etal, Nature 353,
	331 (1991); L.Resvanis, Europhys. News 23, 172 (1992).}
\REF\kamionkowski{M.Kamionkowski, K.Griest, G.Jungman, \& B.Sadoulet,
	Phys. Rev. Lett. 74, 5174 (1995).}
\REF\Nature{Alcock, C., \etal, 1993, Nature, 365, 621.}
\REF\EROS{Aubourg, E., \etal, 1993, Nature, 365, 623;
        Beaulieu J.P., \etal, 1994, preprint.}
\REF\OGLE{Udalski, A., \etal, 1993, Acta Astronomica, 43, 289;
    Udalski, A., \etal, 1994, Acta Astronomica, 44, 165;
    Udalski, A., \etal,  1994, Acta Astronomica, 44, 227;
    Udalski, A., \etal, 1994, ApJ Lett., 436, L103.}
\REF\Pac{\pac, B, 1986, ApJ, 304, 1.}
\REF\DUO{C.~Alard, private communication.}
\REF\PRL{Alcock, \etal, 1995, Phys. Rev. Lett. 74, 2867.}
\REF\LMCone{Alcock, C., \etal, 1995, to appear in ApJ.}
\REF\Bulgeone{Alcock, C., \etal, ApJ 445, 133 (1995).}
\REF\Bennett{ Bennett, D.P. \etal, 1994, Proceedings of the 5th Astrophysics
        Conference in Maryland: Dark Matter.}
\REF\Bulgetwo{Alcock, C., \etal, 1995, in preparation.}
\REF\Stubbs{Stubbs, \etal, 1993, SPIE Proceedings, 1900, 192.}
\REF\bennettetal{D.P.Bennett 1995, in preparation.}
\REF\Schechter{Schechter, P.L., Mateo, M., \& Saha, A., 1994, PASP, 105, 1342.}
\REF\Griesttime{Griest, K. \etal\ 1995, in preparation.}
\REF\Cook{Cook, K., \etal, 1995, Proceedings
        of IAU Colloquium 155: Astrophysical Applications of Stellar Pulsation,
        Cape Town, February 1995, ASP Conference Series, ed. R.Stobie.}
\REF\ourbinary{K.Griest \etal, proceedings of the Pascos/Hopkins Symposium,
	Baltimore, Maryland (World Scientific 1995).}
\REF\Ceph{ C.Alcock, 1995 \etal, AJ, 109, 1653.} %Cepheids
\REF\Griest{Griest, K., 1991, ApJ, 366, 412.}
\REF\Explore{Alcock, \etal, 1995, ApJ 449, 28.}
\REF\Gates{Gates, E.I., Gyuk, G. \& Turner, M.S., 1995, Phys. Rev. Lett., 74,
        3724.}
\REF\Sackett{Sackett, P. \& Gould, A., 1994. ApJ, 419, 648.}
\REF\Evans{Evans, N.W., 1993. MNRAS, 260, 191;
        Evans, N.W., 1994, MNRAS, 267, 333.}
\REF\Evansjijina{Evans, N.W., \& Jijina, J., 1994. MNRAS, 267, L21.}
\REF\Griestbulge{Griest, K., \etal, 1991, ApJ Lett., 372, L79.}
\REF\Pacbulge{\pac, B. 1991, ApJ Lett., 371, L63.}
\REF\Kiraga{Kiraga M., and \pac, B. 1994, ApJ Lett., 430, L101.}
\REF\Gould{C.H.Han \& A.Gould, ApJ 449, 521 (1995).}
\REF\Zhou{Zhou, H.S., Spergel, D.N. \& Rich, R.M., 1994, ApJ Lett., 440, L13.}
\REF\Pacc{\pac, B., \etal, 1994 ApJ Lett., 435, L113.}
\REF\Spergelbar{For example see,
	D.N.Spergel, in ``The Center, Bulge, and Disk of the Milky Way",
	ed. L.Blitz (Kluwer Academic Press, Dordrecht, 1992).}
\REF\gouldparallax{A.Gould, ApJ 444, 556 (1995).}
\REF\gouldsatellite{A.Gould, ApJ Lett. 441, L21 (1995).}
\REF\parallax{C.Alcock \etal, to appear in ApJ Lett. 1995.}
\REF\Mao{Mao, S. \& \pac, B., 1991, ApJ Lett., 374, L37.}
\REF\Loeb{Gould, A. \& Loeb, A., 1992, ApJ, 396, 101.}

%%%%%%%% end of references %%%%%%%%%%%%
%START OF PAPER
\bigskip\bigskip\bigskip
\centerline{\bf  THE NATURE OF THE DARK MATTER} \foot{
Lectures presented at the International School of Physics ``Enrico Fermi"
Course ``Dark
Matter in the Universe", Varenna, 25 July - 4 August, 1995.}
\bigskip
\centerline{KIM GRIEST}
\smallskip
\centerline{Physics Department}
\centerline{University of California, San Diego, La Jolla CA 92093}
\smallskip
\footline={\hss \tenrm \folio \hss}
%\hbox{ }\smallskip
%\centerline{\bf ABSTRACT}
%We review some recent determinations of the amount of dark matter
%on galactic and larger scales, with special attention to the dark matter in
%the Milky Way.  We then briefly review the motivation for and basic physics
%of several dark matter candidates, including
%Machos, Wimps, axions, and neutrinos.
%We then go into more depth for two candidates, the neutralino
%from supersymmetry, and the baryonic Macho candidate.  For Machos we
%Finally we give more in depth review
%give description of the discovery of Machos
%via gravitational microlensing and the interpretation of the results
%with respect to the dark matter problem.
%
\bigskip\bigskip\bigskip\bigskip
\centerline{\bf 1. Introduction}
\smallskip

The dark matter problem has been around for decades, and there is now
consensus that we don't know what the most common material in the Universe
is \rrr{\ashman}.
It is ``seen" only gravitationally, and does not seem to emit or
absorb substantial electromagnetic radiation at any known wavelength.
It dominates the gravitational potential on scales from tiny dwarf
galaxies, to large spiral galaxies like the Milky Way, to large
clusters of galaxies, to the largest scales yet explored.  The
universal average density of dark matter determines the ultimate
fate of the Universe, and it is clear that the amount
and nature of dark matter
stands as one of the major unsolved puzzles in science.

In this series of talks I will first recall the evidence for
dark matter, with emphasis on the dark matter in our own Galaxy.
This overlaps somewhat with Primack's lecture [these proceedings], so
I will be brief.
I then turn to the dark matter candidates and how we might discover
which (if any) of them actually exists.  Then, I will focus in on
two of my favorite candidates, the supersymmetric neutralino Wimp
candidate, and the baryonic Macho candidate.  For the later candidate,
I will go into some detail concerning the one particular experiment
with which I am involved, and present some results showing, that
over a broad range of masses, this candidate has been ruled out as
the primary constituent of the dark matter in our Galaxy.
For the supersymmetric Wimp and especially the neutrino and axion
candidates I will be brief, since there will be talks
by Masiero [these proceedings] on these topics.

\bigskip
\centerline{\bf 2. Physical Evidence for Dark Matter}
\smallskip
Evidence for dark matter (DM) exists on many scales, and it is important to
remember that the dark matter on different scales may be different --
the dark matter in
dwarf spirals may not be the dark matter which contributes $\Omega =1$;
in fact,
the $\Omega=1$ dark matter may not exist.  This consideration is
especially important when
discussing dark matter detection, since detection is done in the Milky Way
and its environs,
and evidence for dark matter outside the Milky Way may not be relevant.
So, let me start with an inventory of dark matter in the Universe.

The cosmological density of dark matter on different scales is quoted using
$\Omega = \rho/\rho_{crit}$,
where $\rho$ is the density of some material averaged over the Universe,
and $\rho_{crit}$ is the critical density.  Most determinations of Omega
are made by measuring the mass-to-light ratio $\Upsilon$ of some system
and then multiplying this by the average luminosity
density of the Universe: $j_0 = 1.9 \pm 0.1 \times 10^8 h^{-1} \lsun/\msun$
[Kirshner, these proceedings].
Here $h = 0.4 - 1$ parameterizes
our uncertainty of the Hubble constant.
There are methods, such as $\Omega_{baryon}$ from big bang nucleosynthesis,
and potential reconstruction from bulk flows, which do not use depend
upon $j_0$, but methods which involve taking an inventory of material
depend upon it.
For example, the mass-to-light ratio in the solar neighborhood
is $\Upsilon \approx 5$, giving $\Omega_{lum} = 0.003h^{-1} = 0.003 - 0.007$.
If the solar neighborhood is typical, the amount of material in
stars, dust and gas is far below the critical value.

\bigskip
\line{\bf 2.1 Spiral Galaxies \hfil}
\smallskip
The most robust evidence for dark matter comes from the rotation
curves of spiral galaxies.  Using 21 cm emission,
the velocities of clouds of neutral hydrogen can be measured as a function
of $r$, the distance from the center of the galaxy.  In almost all cases,
after a rise near $r=0$, the velocities remain constant out as far as
can be measured.  By Newton's law for circular motion $GM(r)/r^2 = v^2/r$,
this implies that the density drops like $r^{-2}$ at large radius and that
the mass $M(r) \propto r$ at large radii.  Once $r$ becomes greater
than the extent of the mass, one expects the velocities to drop $\propto
r^{-1/2}$, but this is not seen, implying that we do not know how large
the extended dark halos around spirals are.  For example, the rotation
curve of NGC3198 \rrr{\binney} implies $\Upsilon > 30h$, or
$\Omega_{halo} > 0.017$.  The large discrepancy between this number
and $\Omega_{lum}$ is seen in many external galaxies and is
the strongest evidence for dark matter.

It is fortunate that the most secure evidence for dark matter
is in spiral galaxies, since
searches for dark matter can be made only in spiral
galaxies;  in fact only in our spiral, the Milky Way.  Unfortunately,
the rotation curve of the Milky Way is not well constrained, with recent
measurements extending only to 15 to 20 kpc, and having differing
amplitudes and shapes \rrr{\blitz,\merrifield}.
This leads to substantial uncertainty in the amount of dark matter
in our Galaxy.
There are other indicators of the mass of the Milky Way.
By studying the motion of dwarf galaxies (especially Leo I at a distance
of 230 kpc) Zaritsky, \etal\ \rrr{\zaritsky}  find a mass of the Milky Way of
$M_{MW} = 1.25^{+ 0.8}_{-0.3} \times 10^{12} \msun$, for $\Upsilon_{MW}
\approx 90$, and $\Omega_{MW} \approx 0.054 h^{-1}$  (assuming the Universe
is like the Milky Way).
A very recent study by Kochanek \rrr{\kochanek}, does a maximum likelihood
analysis including constraints from satellite velocities, the distribution
of high velocity stars (local escape velocity), the rotation curve,
and the tidal effects of M31, to find a mass of the Milky Way inside
50 kpc of $5.4 \pm 1.3 \ten{11} \msun$.  It is interesting that
this value is just what one expects from a flat rotation curve with
$v=220$ km/sec out to 50 kpc, so the Milky Way is very likely a typical
spiral with a large dark halo.

\bigskip
\line{\bf 2.2 Clusters of Galaxies \hfil}
\smallskip
Moving to larger scales, the methods of determining $\Omega$ become
less secure, but give larger values.  There is a great deal of new
evidence on dark matter in clusters of galaxies, coming from
gravitational lensing \rrr{\tyson}
from X-ray gas temperatures \rrr{\briel}
and from the motions of cluster member galaxies.
For example, consider the Coma cluster which contains
around a thousand galaxies.
White \etal\ \rrr{\white}  recently collated some of the data on the Coma
cluster, reporting separate measurements of the amount of mass in
stars, hot gas, and in total.  Within a radius of 1.5$h^{-1}$ Mpc, they
give
$$\eqalign{
M_{star} =& 1.0 \pm 0.2 \times 10^{13} h^{-1} \msun \cr
M_{gas} =& 5.4 \pm 1 \times 10^{13} h^{-5/2} \msun \cr
M_{total} =& 5.7 - 11 \times 10^{14} h^{-1} \msun, \cr}
$$
where the total mass is estimated in two completely different ways.
The first method is a refinement of Zwicky's method of
using the radial velocities of the member galaxies, and the assumption
of virialization to gauge the depth of the gravitational potential well.
The second method makes use of the ROSAT X-ray maps and the assumption of
a constant temperature equilibrium to get the same information.
Remarkably the two methods give the same mass within errors.
Thus with a mass-to-light ratio of $\Upsilon = 330 - 620 \msun/\lsun$,
one finds $\Omega= 0.2 - 0.4$, if the inner 1.5 Mpc of
Coma is representative of the Universe as a whole.

There is, however, a disconcerting in about the above numbers.
As pointed out by White, \etal\ \rrr{\white}
$$
{M_{baryon} \over M_{total}} > 0.009 + 0.05 h^{-3/2}.
$$
Now the Coma cluster is large enough that one might expect its baryon
to dark matter ratio to be the Universal value,
($\Omega_{baryon}/\Omega_{total} = M_{baryon}/M_{total}$), and in fact
White, \etal\ argue that this is the case.  Then the inequality above
should apply to the entire Universe.  But,
big bang nucleosynthesis limits $\Omega_{baryon} < 0.015h^{-2}$.
If $\Omega_{total} = 1$,
the two inequalities are in quite strong disagreement for any value of $h$.
So this is a puzzle.  The conclusions of White, \etal, are that
either $\Omega$ is not unity, or that big bang nucleosynthesis is not
working.
However, there are other possible explanations, notably that
the measurements of the total mass in clusters by gravitational lensing
tend to give larger total mass than the X-ray and virial methods,
and that mass and velocity bias may mean that clusters are not
representative of the Universe as a whole.  The story is
clearly not yet finished.

\bigskip
\line{\bf 2.3 Large Scale Flows \hfil}
\smallskip

It would be best to measure
the amount of dark matter on the largest possible scales so that the
sample is representative of the
entire Universe.
Within the past several years a host of large-scale flow
methods have been
tried and are giving impressive results \rrr{\Dekel}.
These methods have the advantage stated above
but the disadvantage that they
depend upon assumptions about galaxy formation---that is, they depend
upon gravitational instability
theory, the assumption of linear
biasing, etc. Also, the errors in these measurements are
still large and the calculations are
complicated, but they do have great promise, and
tend to give values of $\Omega$ near unity.

A simple example comes from
the observation that the local group of galaxies moves
at $627 \pm 22$ \kmsec\ with respect to the cosmic
microwave background (CMB) (measured from the amplitude of the CMB dipole).
If this motion comes from gravity, then the direction of the motion
should line up with the direction
where there is an excess of mass, and the velocity should be
determined by the size of this excess.
Thus, taking into account the expansion of the Universe, one has
$$
v \propto \Omega^{0.6} {\delta\rho\over\rho} =
     {\Omega^{0.6} \over b}      {\delta n\over n},
$$
where the linear bias factor $b$ has been introduced
to relate the observed excess in galaxy number
counts $\delta n/n$ to  the excess in mass density
$\delta\rho/\rho$.
Using galaxy counts from the IRAS satellite survey,
Yahil \etal\ \rrr{\yahiletal} find
that the direction of the $\delta n/n$ excess agrees with the
direction of the velocity vector to within
$\sim 20^0$, and that
$$
     \beta \equiv{\Omega^{0.6}\over b} = 0.9 \pm 0.2.
$$
Thus with the very conservative limit $b>0.5$, one has
$\Omega>0.2$, and with
the reasonable limit $b>1$, one finds $\Omega>0.5$.
For this method to be reliable, $\delta n/n$ must be measured
on very large scales to ensure that
convergence has been reached, and it is not sure that this is the case.

The above technique is only one of many related methods used to
determine $\Omega$ on large
scales.  Another example is the detailed comparison of the peculiar
velocities of many galaxies with
the detailed maps of $\delta n/n$.  This should not only
determine $\Omega$, but serve as a stringent
test for the theory that large-scale structure is
formed by gravitational instability.
A recent review by Dekel \rrr{\Dekel}
surveys many such methods and concludes that
reasonable evidence exists for $\Omega>0.3$.
Although these techniques holds much promise, it should be noted that
different analyses of the same data sometimes lead to
different conclusions. So for the time being, these
estimates of $\beta$ should not yet be viewed as robust \rrr{\davisnusser}.

In conclusion, the observational evidence for large amounts of dark
matter on galactic halo scales is
overwhelming.  On larger scales, the  observational evidence
for $\Omega$ in the 0.1 to 0.2 range is strong.  On the largest scales,
substantial observational
evidence exists for
$\Omega>0.3$, and some evidence for $\Omega$ near unity exists,
although this may be in conflict
with observations on cluster scales.
\bigskip

\line{\bf 2.4 The Baryonic Content of the Universe \hfil}
\smallskip

An important ingredient in the motivation for non-baryonic dark matter
comes from
big-bang-nucleosynthesis limits on the average baryonic content of
the Universe.
To agree with the measured abundances of helium, deuterium, and
lithium, the baryonic content of the Universe must
be between $0.01 \la \Omega_b h^2 \la 0.015$
\rrr{\bbn,\smithkawmal,\walkeretal}.
Given the large uncertainty in $h$ this means
$0.01 \la \Omega_b \la 0.1$.  These values are far below unity, so
the theoretical predilection for
$\Omega_{total}=1$ (or the
observational evidence for $\Omega \ga 0.3$) forces the bulk of the
dark matter to be non-baryonic.
The lower limit of this range is actually
{\it above} the abundance of known stars, gas, etc., and so there
also seems to be evidence for
substantial {\it baryonic} dark matter as well.

However, if one considered only the most secure dark matter,
that found in spiral galaxies, then it is completely possible that it is
all baryonic.
Since this is the only dark matter which is directly accessible to
experimental detection, it is crucial to consider the possibility
of an entirely baryonic dark halo.

\bigskip
\line{\bf 2.5 Distribution of Dark Matter in the Milky Way \hfil}
\smallskip

While we don't know what the dark matter (DM) is,
we have a fairly reasonable idea as to how much of it there is in the Galaxy,
how it is distributed, and how fast it is moving.
This information comes from the rotation curve of the Milky Way,
and is crucial to all the direct searches for dark matter.
If we say that the rotation curve of the Milky Way is constant at about
$v_c = 220$ km/sec out to as far as it is measured, then we know that
the density must drop as $r^{-2}$ at large distances.
This velocity also sets the scale for the depth of the potential well
and says that the dark matter must also move with velocities in this range.
Assuming a spherical and isotropic velocity distribution is common,
and a usual parameterization is
$$
\rho({\bf r}) = \rho_0 {a^2 + r_0^2\over a^2 + r^2},
$$
where $r_0 \approx 8.5$ kpc is the distance of the Sun from the galactic
center, $a$ is the core radius of the halo, and $\rho_0 \approx 0.3\
{\rm GeV~cm}^{-3}$ is the density of dark matter near the Sun.
Also, a typical velocity distribution is
$$
f(v) d^3v = {e^{-v^2/v_c^2} \over \pi^{3/2}v_c^3} d^3v.
$$
It should be noted that the specifics of the above models are not
very secure.  For example, it is quite possible that the halo of
our Galaxy is flattened into an ellipsoid, and there may be a component of
the halo velocity which is rotational and not isotropic.
Also, some (or even most) of the rotation curve of the Milky Way
at the solar radius could be due to the stellar disk.
Canonical models of the
disk have the disk contributing about half the rotation velocity,
but larger disks have been envisioned.  Recent microlensing results
may be indicative of a larger disk as well (see Section 7.).

Finally, other important points about our Galaxy's geography include
the fact that the nearest two galaxies are the LMC and SMC, located
at a distance of 50 kpc and 60 kpc respectively, that the halo of the
Milky Way is thought to extend out at least this far, and that
the bulge of the Milky Way is a concentration of stars in the center
of our Galaxy (8.5 kpc away) with a size of about 1 kpc.

\bigskip
\centerline{\bf 3. Brief Survey of Dark Matter Candidates}
\smallskip
There is no shortage of ideas as to what the dark matter could be.
In fact, the problem is the opposite.  Serious candidates
have been proposed with masses ranging
from $10^{-5}$ eV = $1.8 \times 10^{-41}$ kg $= 9 \times 10^{-72} \msun$
(axions) up to $10^4 \msun$ black holes.  That's a range of masses
of over 75 orders of magnitude!  It should be clear that no one
search technique could be used for all dark matter candidates.

Even finding a consistent categorization scheme is difficult, so
we will try a few.  First, as discussed above, is the baryonic
vs non-baryonic distinction.  The main baryonic candidates
are the Massive Compact Halo Object (Macho) class of candidates.
These include
brown dwarf stars, jupiters, and 100 $\msun$ black holes.
Brown dwarfs are spheres of H and He with masses below 0.08 $\msun$,
so they never begin nuclear fusion of hydrogen.
Jupiters are similar but with masses near 0.001 $\msun$.
Black holes with masses near 100 $\msun$
could be the remnants of an early generation of stars
which were massive enough so that not many heavy elements were
dispersed when they underwent their supernova explosions.
Other, less popular, baryonic
possibilities
include fractal or specially placed clouds of molecular
hydrogen \rrr{\depaolis}.
The non-baryonic candidates are basically elementary particles which
are either not yet discovered or have non-standard properties.
Outside the baryonic/non-baryonic categories
are two other possibilities which
don't get much attention, but which I think should be kept in mind
until the nature of the dark matter is discovered.
The first is non-Newtonian
gravity.  See Begeman \etal\  \rrr{\mond} for a provocative discussion of this
possibility; but watch for results from gravitational lensing which
may place very strong constraints.
Second, we shouldn't ignore the ``none-of-the-above" possibility
which has surprised the Physics/Astronomy community several times
in the past.

Among the non-baryonic candidates there are several classes of
particles which are distinguished by how
they came to exist in large quantity
during the Early Universe,
and also how they are most easily detected.
The axion (Section 5) is motivated as a possible solution
to the strong CP problem and is in a class by itself.  The largest
class is the Weakly Interacting Massive Particle (Wimp)
class (Sections 4 and 6),
which consists of literally hundreds of suggested particles.
The most popular of these Wimps
is the neutralino from supersymmetry (Section 6).
Finally, if the tau and/or muon neutrino had a mass in the 2 eV to 100 eV
range, they could make up all or a portion of the dark matter.
This possibility will be discussed by Masiero and also Klypin
[these proceedings].

Another important categorization scheme is the ``hot" vs ``cold"
classification.  A dark matter candidate is called ``hot" if it was moving
at relativistic speeds at the time when galaxies could just start to form
(when the horizon first contained about $10^{12}\msun$).  It is called
``cold" if it was moving non-relativistically at that time.
This categorization
has important ramifications for structure formation, and there is
a chance of determining whether the dark matter is hot or cold
from studies of galaxy formation.  Hot dark matter cannot cluster
on galaxy scales until it has cooled to non-relativistic speeds,
and so gives rise to a considerably different primordial fluctuation
spectrum [see Klypin, these proceedings].
Of the above candidates only the light neutrinos would be hot;  all the
others would be cold.
\bigskip

\centerline{\bf 4. Thermal Relics as Dark Matter (Wimps) \hfil}
\smallskip

Among the particle dark matter candidates an important distinction
is whether the particles were created thermally in the Early Universe,
or whether they were created non-thermally in a phase transition.
Thermal and non-thermal relics have a different relationship between
their relic abundance $\Omega$ and their properties
such as mass and couplings, so the distinction is especially important
for dark matter detection efforts.
For example, the Wimp class of particles can be defined as those particles
which are created thermally, while dark matter
axions come mostly from non-thermal processes.

In thermal creation one imagines that early on, when the Universe
was at very high temperature, thermal equilibrium obtained, and
the number density of Wimps (or any other particle species)
was roughly equal to the number density of photons.
As the Universe cooled the number of Wimps and photons would decrease together
as long as the temperature remained higher than the Wimp mass.
When the temperature finally dropped below the Wimp mass, creation
of Wimps would require being on the tail of the thermal distribution,
so in equilibrium, the number density of Wimps would drop
exponentially $ \propto \exp(-m_{Wimp}/T)$.  If equilibrium were maintained
until today there would be very few Wimps left, but at some point
the Wimp density would drop low enough that
the probability of one Wimp finding another
to annihilate would become small.  (Remember we must assume that an individual
Wimp is stable if it is to become the dark matter.)
The Wimp number density would ``freeze-out" at this point
and we would be left
with a substantial number of Wimps today.  Detailed evolution
of the Boltzmann equation can be done for an accurate
prediction [Section 6.2], but roughly
$$
\Omega_{Wimp} \approx {10^{-26} {\rm cm}^3 {\rm s}^{-1}
\over \VEV{\sigma v}},
$$
where $\VEV{\sigma v}$ is the thermally averaged cross section
for two Wimps to annihilate into ordinary particles.
The remarkable fact is that for $\Omega \approx 1$, as required by
the dark matter problem, the annihilation cross section
$\left<\sigma v\right>$ for any thermally created particle
turns out to be just
what would be predicted for particles with electroweak scale interactions.
Thus the ``W" in ``Wimp".  There are several theoretical problems with the
Standard Model of particle physics which are solved by new electroweak
scale physics such as supersymmetry.  Thus these
theoretical problems may be clues that the dark
matter does indeed consist of Wimps.  Said another way, any stable
particle which annihilates with an electroweak scale cross section
is bound to contribute to the dark matter of the Universe.  It is
interesting that theories such as supersymmetry, invented for entirely
different reasons, typically predict just such a particle.

The fact that thermally created dark matter has weak scale interactions
also means that it may be within reach of accelerators such as LEP at CERN,
and CDF at Fermilab.  After all these accelerators were built
precisely to probe
the electroweak scale.  Thus many accelerator searches for exotic particles
are also searches for the dark matter of the Universe.
Also, due to the weak scale interactions, Wimp-nuclear
interaction rates are within reach of many direct and indirect
detection methods (see Section 6).

\bigskip
\centerline{\bf 5. Non-thermal Relics as Dark Matter (Axions) \hfil}
\smallskip

\def\tbar{{\bar\theta}}
The best example of a non-thermal particle dark matter candidate
is the axion \rrr{\axion}.
Actually, thermal axions are produced in the standard
way, but if such axions existed in numbers so as to make up the dark
matter, they would have lifetimes too short to still be around in
quantity.
However, there is another, more important, production mechanism
for axions in the early Universe.

The axion arises because the QCD Lagrangian contains a term
$$ L \supset {\tbar g^2 \over 32 \pi^2} G \tilde G,
$$
where $G$ is the gluon field strength.
This term predicts an electric dipole moment of the neutron of
is $d_n \approx 5 \times 10^{-16} \tbar$.
Experimentally, however, the neutron dipole moment
$d_n < 10^{-25}$, which means $\tbar  < 10^{-10}$.
The question is why does this $\tbar$ parameter have such
a small value, when it naturally would have a value near unity?
This is the strong CP problem, and one way to resolve this
problem is to introduce a new Peccei-Quinn symmetry which predicts
a new particle -- the axion.  The P-Q symmetry forces $\tbar=0$
at low temperatures today, but in the early Universe,
the axion field was free to roll around the bottom
of its Mexican hat potential.  The axion field motion in
the angular direction is called $\theta$, and since the curvature of
the potential in this direction is zero, the axion at high temperatures
was massless.  However, when the temperature of the Universe cooled
below a few hundred MeV (QCD energy scale), the axion potential
``tilts" due to QCD instanton effects, and the axion begins to oscillate
around the minimum, like a marble in the rim of a tilted Mexican hat.
The minimum of the potential forces the average $\tbar$ to zero,
solving the strong CP problem, and the curvature of the potential
means the axion now has a mass.  There is no damping mechanism for
the axion oscillations, so the energy density which goes into oscillation
remains until today as a coherent axion field condensate filling the
Universe.  This is a zero momentum condensate and so constitutes
cold dark matter.  One can identify this energy density as a bunch
of axion particles, which later can become the dark matter
in halos of galaxies.  The relic energy density $\Omega$ is thus related
to the tilt of the potential, which in turn is related to the axion
mass, a free parameter of the model.  If the axion mass
$m_a \approx 10^{-5}$ eV, then $\Omega_a \approx 1$.
One now sees why axions are cold dark matter even though they are so light.
This rather unusual story is probably the most elegant
solution to the strong CP problem, and several groups are
mounting laboratory searches for the coherent axions which may make up
the major component of mass in the Galaxy.
For example, a group involving physicists from Lawrence Livermore National
Lab, the Russian INR, the University of Florida, MIT, Fermilab,
UC Berkeley, LBL, and the University of Chicago \rrr{\vanbibber}
is building an
extremely loise noise microwave cavity inside of a large magnetic
field for this purpose.  The basic idea is that halo axions can
interact with the magnetic field and produce microwave photons.
This will happen resonantly when the cavity is tuned precisely to the
axion mass, so one scans the frequency spectrum looking for such
a resonance signal.  Two experiments, one at Florida and one at
Brookhaven have already used this technique and reported negative
results \rrr{\hagmann}.
The sensitivity of thoses early experiments was significantly below
the expected signal, however, and it is this new experiment which
will for the first time have the capability of detecting dark matter
axions if in fact they exist.
\bigskip

\centerline{\bf 6. Search for Wimp Dark Matter (Neutralinos)}
\smallskip

Why is it important to actually search for and to identify the dark matter?
Of course it is intrinsically interesting to know what the primary constituent
of the Universe consists of, but also until we know the dark matter identity,
there will always be the doubt that there is no dark matter, and instead
there is some flaw in our knowledge of fundamental physics.

Unfortunately, no one technique is useful for all the different candidates.
The only way to proceed is to pick a candidate
and design an experiment specific to that candidate.  This is risky
proposition, especially for the experimentalist who must spend many
years of his or her life developing the technology and performing the
search.  For this reason, only the best motivated candidates are currently
being searched for.  There are hundreds of other dark matter candidates
that we have not discussed at all.
As one goes down the list of popular candidates,
asking oneself which candidate is the most likely,
I have to admit that ``none-of-the-above"  comes to mind.  However,
it is difficult to make any progress searching for an unspecified
candidate.  After ``none-of-the-above", I think the Wimp candidates,
and especially the supersymmetric neutralino candidate, is probably the
best bet.  It may sound odd to an astrophysically oriented group such
as yourselves that my best guess for the dark matter is a specific
undiscovered elementary particle based on a theory for which there is
no evidence, so I will spend some time describing my somewhat
idiosyncratic reasons.

Please see the lectures by A.~Masiero [these proceedings] for
additional in depth discussion of many of the issues covered here.
Also see the new Physics Report ``Supersymmetric Dark Matter", by
Jungman, Kamionkowski, and myself for more details on everything covered
here.
\bigskip

\line{\bf  6.1 Motivation for Supersymmetry \hfil}
\smallskip

First, why should astrophysicists take seriously supersymmetry, a theory
which requires more than doubling the number of known
elementary particles, none of which has yet been detected?
When Dirac attempted to make special relativity consistent with
quantum mechanics he discovered the Dirac equation.  He also
discover a disconcerting fact.
There was a new symmetry, CPT symmetry, implied by his equation, and
this symmetry required that for every known particle there had exist a
charge conjugate, or anti-particle.  He resisted the idea of doubling
the number of known particles, and initially hypothesized that the CPT
partner of the electron was the proton.  This idea was soon shown to
be impossible, but fortunately the anti-electron and anti-proton were
soon discovered, vindicating Dirac's theory.

The situation may be similar with regard to supersymmetry.  Many attempts
have been made to make general relativity consistent with quantum field
theory, especially within the framework of a theory which combines
gravity with the strong and electroweak interactions.  It is interesting
that in all the most successful attempts a new symmetry is required.
The powerful Coleman-Mandula theorem states that within the framework
of Lie algebras, there is no way to unify
gravity with the gauge symmetries which describe the strong and electroweak
interactions. So the ``super"-symmetry which successfully combines these
interactions had to move beyond Lie algebras to ``graded" Lie algebras.
Graded Lie algebras are just like Lie algebras except they use anti-commutation
relations instead of commutation relations.  Thus they relate particles
with spin particles to without spin.
Examples of theories that attempt to combine gravity
with the other forces include super-strings and super-gravity,
where in both cases ``super" refers to the supersymmetry.
Thus, if such a symmetry exists in nature, every particle with spin (fermion)
must have a related super-symmetric partner without spin (boson),
and vice versa.
As it now stands, standard quantum field theory
seems to be incompatible with general relativity.  Since the world
is unlikely to be incompatible with itself, it seems
either quantum field theory or general relativity
must be modified, and of course a new theory which
combines gravity with the strong and electroweak interactions would be
the most elegant.  Thus, one sees why so many particle physicists
have become enamored with supersymmetry, and why many thousands of papers
have been written on the subject.

As in Dirac's case, this doubling of the number of particles was disconcerting,
and it was initially hoped that perhaps the neutrino could be supersymmetric
partner of the photon.  Now it is known that this is impossible,
but unlike in Dirac's case, no discovery of supersymmetric partners has
quickly followed.  In fact, it is now known that supersymmetry must be
a ``broken" symmetry, since perfect supersymmetry requires that
the masses of the super-partners be the same as their counterparts.
This is easily arranged, but leaves the masses of all the superpartners
undetermined.  In fact, the masses could be so large that all the
superpartners are completely undetectable in current or future accelerators
and are therefore mostly irrelevant to current physics or dark matter
detection.  There are however some very suggestive reasons why the
superpartners may have masses in the 100 GeV to several TeV range.

First, there is coupling constant unification.  The strength of the strong,
weak, and electromagnetic interactions is set by the value
their coupling constants, and these ``constants" change as the energy of the
interactions increase.  For example, the electromagnetic coupling constant
$\alpha = {1\over 137}$, has a value near $1\over 128$ when electrons
are collided at the LEP machine at CERN.  Several decades ago it was
noticed that the three coupling constants would meet together at
a universal value when the energy of interactions reached about $10^{15}$
GeV.  This would allow a ``Grand Unification" of the strong, weak, and
electromagnetic interactions, and much model building was done.
In the past few years, the values of the three coupling constants have
been measured much more accurately, and it is now clear that, in fact,
they cannot unify at any scale unless many new particles are added to
the theory.  Suggestively, if the supersymmetric partners exist,
and have reasonably low masses they give just the right contribution
to force the coupling constants to unify.

Next, there is the gauge hierarchy  problem.
The standard model of particle physics is enormously successful.
It accurately predicts the results of hundreds of measurements.
In the standard model,
fermions such as electrons are intrinsically massless, but develop
a mass through interactions with the Higgs field that is hypothesized to
fill the Universe.  The mass of a fermion then is just proportional
to the strength of its coupling to the Higgs field.  The Higgs is
thus an essential feature of the standard model.  The Higgs also
develops a mass through a ``bare" mass term and interactions with other
particles, but due to is scalar nature, the mass it acquires through
interactions are as large as the largest mass scale in the theory.
Thus, in any unified model, the Higgs mass tends to be enormous.
Such a large Higgs mass cannot be, however, since it would ruin the
successful perturbation expansion used in all standard model calculations.
Thus in order to get the required low Higgs mass,
the bare mass must be fine-tuned to dozens of significant places
in order to precisely cancel the very large interaction terms.  At
each order of the perturbation expansion (loop-expansion), the procedure
must be repeated.
However, if supersymmetric partners are included, this fine-tuning
is not needed.  The contribution of each supersymmetric partner cancels off
the contribution of each ordinary particle.  This works only if the
supersymmetric partners have masses below the TeV range.  Thus, stabilization
of the gauge hierarchy is accomplished automatically, as long as
supersymmetric particles exist and have masses in the range 100 -1000 GeV.
The enormous effort going into searches for supersymmetric particles at
CERN, Fermilab, etc.  is largely motivated by this argument.

Even though no supersymmetric particles have been discovered, they
have all been given names.  They are named after their partners.
Bosonic ordinary particles have fermonic superpartners with the same name
except with the suffix
``ino" added, while fermonic ordinary particles have bosonic (scalar)
superpartner names with the prefix ``s" added.  So for example, the photino,
Higgsino, Z-ino, and gluino are the partners of the photon, Higgs, Z-boson,
and gluon
respectively.  And the squark, sneutrino, and selectron are the scalar
superpartners of the quark, neutrino, and electron respectively.
There are several superpartners which have the same quantum
numbers and so can mix together in linear combinations.  Since these do
not necessarily correspond to any one ordinary particle, they are given
different names.  For example, the photino, Higgsino, and Z-ino can
mix into arbitrary combinations called the neutralinos, and the
charged W-ino and charged Higgsino combine into particles called
charginos.

Finally, an interesting feature of most supersymmetric models is the
existence of a multiplicatively conserved quantum number called R-parity,
in which each superpartner is assigned $R=-1$, and each ordinary particle
is assigned $R=+1$.  This quantum number implies that supersymmetric particles
must be created or destroyed in pairs, and that the lightest supersymmetric
particle (LSP) is absolutely stable;  just as the electron is stable
since electric charge is conserved and there is no lighter charged
particle into which it could decay.
This fact is what makes supersymmetric particles
dark matter candidates.
If supersymmetry exists and R-parity is conserved, then some LSP's
must exist from the Early Universe.
The only question is how many.

\bigskip
\line{\bf 6.2 Relic Abundance in More Detail \hfil}
\smallskip

The number density of any particle which was once in thermal equilibrium
in the Early Universe can be found by solving the relevant set of
Boltzmann equations.  In most cases only one is needed
\def\nneut{n_\chi}
\def\nstuff{n_{ord}}
\def\sigmav{\langle \sigma v \rangle}
$$
{d\nneut\over dt} = - 3H\nneut - \nneut^2 \sigmav(\chi \chi\rightarrow
{\rm ordinary\ stuff}) + \nstuff^2 \sigmav({\rm ordinary\ stuff} \rightarrow
\chi\chi),
$$
where $H$ is the Hubble constant, $n$ is the number density, $t$ is time, and
$\sigmav$ is the thermally averaged cross section times the relative
velocity of the interacting particles.
We are using $\chi$ to denote the LSP.
The first term on the right-hand
side is the reduction in number LSP density due to the Hubble expansion,
the second term is reduction due to self-annihilation, and the third
term is the increase due to particle production.  The third term
can be simplified using the fact that ``ordinary" particles such as
quarks and electrons stay in thermal equilibrium throughout period
during which the Wimp number density ``freezes out" (see Section 4).
When thermal
equilibrium obtains, creation equals annihilation, so
the second and third terms are equal.  Therefore one can eliminate
the ``ordinary particle" cross sections and number densities and find
the usual equation
$$
{dn\over dt} = - 3H\nneut - (\nneut^2 -{\nneut^{eq}}^2)
\sigmav_{annihilation}.
$$
Starting at an early time when all particles were in equilibrium,
one integrates this
equation either numerically or using
the standard ``freeze-out" approximation \rrr{\kolb}.
and obtains the number density at $t=0$ (today).  The relic abundance
is simply $\Omega_\chi = m_\chi \nneut /\rho_{crit}$, where $m_\chi$
is the mass of the LSP.

The difficult step in obtaining the current day density, is usually
the calculation of the annihilation cross section of two LSPs
into all standard model particles.  In order to perform this calculation,
one must first determine which particle is the LSP, and then evaluate
all the relevant Feynman diagrams.
Going through the list of supersymmetric particles, one finds, basically
by process of elimination, that only the sneutrino and neutralino
are likely candidates.  In the vast majority of models,
the neutralino is favored
over the sneutrino, so most work has concentrated on the neutralino
as dark matter candidate.

For the neutralino, several dozen Feynman diagrams contribute
to self-annihilation,
including possible annihilation into quarks, leptons, W, Z, and Higgs bosons,
and involving most of the super-partners as exchange particles.  So in order
to perform the calculation one needs to first obtain the mass and couplings
of all the supersymmetric particles.  Since supersymmetry is broken,
the mass terms are unknown, giving rise to many free parameters in the
most general supersymmetric model.  Usually, in order to simplify things,
one considers the ``minimal" supersymmetric model, the model with the fewest
number of new particles, but still there are many undetermined parameters.
So to further simplify, several other assumptions are usually made.
In minimal supergravity, some GUT scale assumptions can reduce the number
of parameters to just a few.  In what follows, we use some, but not
all, of the super-gravity assumptions, and
have as a result 5 free parameters \rrr{\Jungman}.
The parameters are the gaugino mass parameters $M_1$ and $M_2$,
the Higgsino mass parameter $\mu$, the pseudoscalar Higgs mass $m_A$,
and the ratio of Higgs vacuum expectation values $\tan\beta$.
For any set of parameters, one can calculate all the masses, mixings,
\def\omh{\Omega h^2}
and couplings, and then the annihilation cross section.  Thus after
a long calculation, one finally obtains $\omh$ in terms of the five
parameters.  If one obtains a relic abundance in the range
$0.025 < \omh < 1.0$, then that set of parameters defines a potential
dark matter candidate.  However, before deciding that this is a
dark matter candidate one must ensure that one of the many accelerator
experiments that have searched for supersymmetric particles has
not already ruled out that model.
\bigskip

\line{\bf 6.3 Accelerator Constraints \hfil}
\smallskip

Extensive unsuccessful searches for the particles involved in supersymmetric
models have been performed at particle accelerators throughout the world.
This does not yet mean that low-energy supersymmetry is unlikely to exist
since only a small portion of the mass range under 1 TeV has been explored.
However, substantial regions of prime neutralino dark matter parameter space
have been eliminated, and it is important to check this when considering
the detectability of any neutralino candidate.  One does not want to
build a detector only capable of seeing particles ruled out by current
experiments.  In the following, we demonstrate a method of exploring
supersymmetric parameter space taking into account
accelerator constraints in a rough way \rrr{\Jungman}.
Note that the same supersymmetric parameters which determine the relic
abundance cross sections determine all the particle production and
rare decay cross sections.  Thus once these parameters are specified,
one can compare the model predictions with experimental results.
A partial list of relevant experimental results follows.
Higgs searches at LEP rule out the lightest scalar Higgs masses below
about 45 GeV, and pseudoscalar Higgs masses below about 39 GeV, using
cross sections such as $Z\rightarrow h\mu^+\mu-$, and
$Z\rightarrow h A$.  LEP chargino searches at the Z pole rule
\def\mcharg{m_\chi^\pm}
out $\mcharg$ below 45 GeV, and direct neutralino searches constrain
the branching ratio of Z into neutralinos to be less than about $10^{-5}$.
The squark and gluino searches by CDF give complicated results, but one
is probably safe if one limits consideration to squarks with mass larger
than 150 GeV.  Finally, the recent CLEO measurement of
$10^{-4} < {\rm BR}(b\rightarrow s \gamma) < 4.2 \ten{-4}$
has important consequences for neutralinos.  This is the decay of
bottom quarks into strange quarks plus a photon, and the measurement
is within the prediction of the standard model.  The impact on supersymmetry
comes because this process can also occur via exchange of supersymmetric
particles and in many cases these contributions can destroy the experimental
agreement with the standard model.   So this branching ratio should also
be computed for every set of supersymmetric parameters, and models which
do not agree with the above constraint should be eliminated.
We illustrate the process by considering the grid of models in Figure~1.
Since the actual parameter space is five-dimensional, this is just a two-
dimensional projection of the parameters.
Figure~1(a) shows the entire grid of models, while Figure~1(b) shows
the models which are left after eliminating those which violate an
accelerator constraint (or other consistency test).

Using just the allowed models we can not plot the neutralino mass vs
the relic abundance.  The resulting plot (Figure~2)
is quite remarkable and can be
taken as a hint that supersymmetry may well have something to do with
the dark matter problem.  Many models fall in the $0.025 < \omh < 1.0$
range.  Recall that models with $\omh>1$ imply a dark matter density
inconsistent with cosmological measurements.  Thus dark matter considerations
can be used to help the particle physicists in their search for supersymmetry;
there is probably little use in considering models which are inconsistent
with cosmology (though as experimentalists, it is probably wise that not
too much weight is given to such results).  On the other hand, models
with $\omh< 0.025$ are perfectly viable from a particle physics point
of view, but predict too little relic abundance to make up all of the
dark matter.  It is interesting to note, however, that even a relic
abundance of $\Omega = 10^{-5}$ would make neutralinos as large a
contributor as the microwave background.  There was no fine tuning
invoked to produce the numerous models with relic abundance in the proper
range to be the dark matter, and it seems that no matter what, if
stable neutralinos exist, they must be an important contributor to
the mass inventory of the Universe.

\bigskip
\line{\bf 6.4 Detection techniques \hfil}
\smallskip

There are several ways of attempting to test the hypothesis that stable
neutralinos exist and contribute to the dark matter.  History has shown
that the most powerful method of discovering new particles is with particle
accelerators, so if I had to guess, I would guess that
discovery of supersymmetric dark matter will come from CERN.  The
new LEP 200 machine should be coming on line in a few years, and it
has the ability to explore much of the minimal supersymmetric parameter
space.  The most powerful search will be their Higgs search, and if they
find a Higgs which is not the standard model Higgs, I would take it as
strong evidence for supersymmetry.  New searches for neutralinos and other
supersymmetric partners will also be made, so anyone interested in the
identity of the dark matter should watch for these results.
After LEP 200, the cancelled SSC, had the best chance of discovering
supersymmetry, so that cancellation was a big disappointment.
Luckily, Europe has picked up the ball and the LHC at CERN has now
been funded to search for the Higgs and supersymmetry.
Keep in mind that if neutralinos are discovered, and their mass and couplings
measured, one could predict the relic abundance using the methods discussed
above, and know what contribution they make to the dark matter.

While the accelerators have perhaps the best chance of discovering
supersymmetric dark matter, it would be much more satisfying to actually
detect the particles in our halo as they move past and through the Earth.
This would also allow measurement of the local density of dark matter
and establish beyond doubt that the dark matter is non-baryonic cold
dark matter.  Currently there are two main methods being aggressively
pursued.

\bigskip
\line{\bf 6.5 Direct Detection \hfil}
\smallskip

The most exciting result would be direct detection of the Wimp particles.
Since we roughly know the speed ($\sim 270$ km/s) and the density
($\rho \sim 0.3$ GeV cm$^{-3}$), we can say that for a Wimp of mass
of order 10-100 GeV, roughly 100,000 dark matter particles a second pass
through every square centimeter of the Earth, including our bodies.
If they exist, these are very weakly interacting particles, so it is
quite rare that one of them will interact at all.
In addition, if one does elastically scatter off a nucleus, the deposited
energy is usually in the keV to 100 keV range, so extremely sensitive
devices must be used.  These difficulties, however, have not stopped many
groups throughout the world from developing devices capable of detecting
Wimps.  See references \rrr{\Jungman,\direct} for details.

In deciding the size, sensitivity, and energy threshold of
a detector, the experimentalist would like to know what event rate one
expects in the case that the dark halo consists entirely of Wimps.  For
a unspecified Wimp, only rough estimates can be made using
general arguments \rrr{\griestsadoulet}, but for neutralino
Wimps, the elastic scattering cross section and the event rate
per kilogram detector per year can be calculated, once the supersymmetric
model parameters are chosen.  Figure~3 shows a scatter plot of the
rate in a germanium 73 detector, for all the models that
pass the accelerator constraints and have relic abundances in the range
$0.025 < \omh < 1$.
The ``stripes" in the plot are due to the
finite grid we sampled in parameter space, and so the spaces between the
stripes should be mentally filled in.  A kilogram of germanium was chosen
since this is roughly the material and size of one of the most advanced
experimental efforts (see below).  We see that if neutralinos of around 50 GeV
mass make up the dark matter, the expected event rate is probably between
$10^{-4}$ and 1 event/kg/year.

When a Wimp scatters off a nucleus, the nucleus recoils, causing
dislocation in the crystal structure,
vibrations of the crystal lattice (i.e. phonons or heat), and also
ionization.  The main difficulties in these experiments come from the fact
that the events are rare and that there are many backgrounds which
deposit similar amounts of energy on much more frequent time-scales.  So
in the past few years the main experimental efforts have gone toward
increasing the mass of the detectors and discriminating the nuclear
recoil signal from the background.  Generally the detectors must
be operated deep underground at milli-Kelvin temperatures, and be heavily
shielded.
A illustration of the problem is shown in Figure~4, which shows
the background in a germanium detector built by the Berkeley, LBL,
UCSB group and operated under the Oroville dam \rrr{\angela,\caldwell}.
One sees many background
processes including lines from radioactive elements and tritium,
electron noise at low energy deposited, and a roughly constant
background at about one event/kg/day/keV.  Comparing this to a typical
expected signal in Figure~5, one sees the problem.  However, the vast
majority of the background comes from gamma rays, while the Wimp signal
would be nuclear recoils, and it has been established that gamma rays
deposit a much larger fraction of their energy in ionization than
in phonons or heat.  So the experimentalists measure simultaneously
the energy deposited in heat and the energy deposited in ionization
and are therefore able to reject perhaps 99\% of the background gamma
rays \rrr{\CDMS}.
This kind of discrimination is possible only in materials such as
germanium and silicon which can be used as ionization detectors, but
for other materials such as NaI, and Xenon other effects such as
pulse shape or scintillation light may be used to separate the gamma-rays
from the nuclear recoils \rrr{\otherdirect}.
Using the CDMS (Berkeley/LBL/UCSB/Stanford/Baksan) collaboration as
an example \rrr{\CDMS},
the sensitivity of experiments starting to run this year
is in the 0.1 to 1 event/kg/year range, and upgraded versions hope
to reach 0.01 event/kg/year within a few years.  Returning to Figure~3,
one sees that there are viable supersymmetric models which will be explored
and that a discovery is possible.  However, one also sees that a definitive
experiment will not be possible within the next few years, since rates
below the expected experimental sensitivity are common.  However, it
is remarkable to realize that these small underground experiments are
competing directly with CERN in the race to discover supersymmetry.
And the enormous increases in sensitivity these experiments have
accomplished in the past few years, leads one to expect further such
advances in the future.

\bigskip
\line{\bf 6.6 Indirect Detection \hfil}
\smallskip

A great deal of theoretical and experimental effort has gone into another
potential technique for Wimp detection.  The idea is that if the
halo is made of Wimps, then these Wimps will have been passing through
the Earth and Sun for several billion years.  Since Wimps will occasionally
elastically scatter off nuclei in the Sun and Earth, they will occasionally
lose enough energy, or change their direction of motion enough, to become
gravitationally captured by the Sun or Earth.  The orbits of such captured
Wimps will repeatedly intersect the Sun (or Earth) resulting in the eventual
settling of the Wimps into the core.  As the number density increases
over time, the self-annihilation rate $\chi\chi \rightarrow
\nu {\bar\nu}$ will increase, resulting in a stream of neutrinos
produced in the core of the Sun or Earth.  Neutrinos easily escape the Solar
core and detectors on Earth capable of detecting neutrinos coming from
Sun or Earth have operated for some time.  The energy of such neutrinos is
roughly 1/2 to 1/3 the mass of the Wimp, so these neutrinos are much
higher energy than the MeV scale Solar neutrinos from nuclear
reactions that have already been detected.  The higher energy
of these Wimp annihilation neutrinos make them easier to detect
than ordinary solar neutrinos and somewhat compensates for their
much fewer numbers.   It also makes them impossible to confuse
with ordinary solar neutrinos.   Thus the presence of a source of
high energy neutrinos emanating from the centers of the Sun and Earth
would be taken as evidence for Wimp dark matter.
While the above chain of reasoning may seem long, I don't know of any
holes in it, and several experimental groups are in the process of
designing and building detectors capable of seeing such a neutrino
signal.
For example, the IMB and Kamionkande proton decay detectors have already
been used to set (very weak) limits on Wimp dark matter using this
technique \rrr{\indirect}.
The MACRO monopole search detector has also looked for
this signal \rrr{\indirect}.
Several new detectors are being created which should be
substantially more sensitive.
For this signal, it is not the mass of the detector
which is relevant, but the surface area.  Neutrinos from the core of
the Sun or Earth produce muons in the atmosphere and rock around the
detectors, and it is primarily these muons the detectors watch for.
Muons are also copiously created by cosmic rays entering the Earth's
atmosphere, so there is a substantial background of  ``downward"
traveling muons.  These detectors, then are located deep underground,
where the rock shields many of the background muons, and they also
focus on ``upward" traveling muons, which are much more likely to
have been created by neutrinos that have traveled through Earth and
interacted in the rock just below the detector.  Thus surprisingly,
the best way to see high energy neutrinos from the Sun is to go deep
underground at night (when the Sun is ``under" the Earth)!
Since the range of the muons depends mostly on the energy of the neutrinos,
The number of muons detected depends mostly on the surface area of the
detector.  So the new generation of detectors are designed to have very
large surface areas.   Examples include MACRO, superkamiokande, AMANDA,
DUMAND, and NESTOR \rrr{\newindirect}.

As an example, consider the AMANDA detector \rrr{\newindirect}
which is being prototyped
in Antarctica.  There are several ways to detect high energy muons, one
of which is to measure the Cerenkov light emitted as they travel
faster than light-speed in some medium.  AMANDA places strings of
phototubes deep in the Antarctic ice, in order to detect the Cerenkov
light thus emitted.  So far four long strings have been deployed
at depths in the kilometer range.  These deep holes are dug in a
day using just hot water!  The Antarctic ice is extremely clear
and light can travel large distances.  Small lasers were also
put down in order to measure the ice transparency and test the feasibility
of the idea.  The initial results were both bad and good.  The collaboration
found bubbles in the ice substantially larger than the ``ice experts"
had indicated.  These meant that the Cerenkov light diffused too much
to be useful in detecting muons.  However, the size of the bubbles is
decreasing with depth, and they expect by placing their next phototube
strings deeper the bubble problem will disappear.  The good news was
that the ice was substantially more transparent that they had expected,
meaning that they can place their next strings further apart, thereby
increasing the effective surface area of their detector.

How will detectors such as AMANDA fare in the detection of Wimps?
Using the cross sections, etc. calculated from the supersymmetric models
one can calculate the density of neutralinos in the Sun and Earth,
and then the annihilation rate, and then the number of neutrinos
incident on Earth, and then the number of muons produced, and finally
the number of muons detected.  An example of such a calculation
is shown in Figure~6, for precisely the same models shown in Figure~3.
The AMANDA detector may have an effective area of 1000 m$^2$, so
as you can see the story is somewhat the same as for direct detection.
There is a region of supersymmetric parameter space which will be
probed by these indirect detectors, but there are many possible sets
of model parameters for which indirect detection is not possible without
much more sensitive detectors.   A comparison of direct and
indirect detection methods leaves one with the impression that for a
typical neutralino a
kilogram of direct detector germanium has about the same sensitivity
as $10^4-10^6$ m$^2$ of indirect detector \rrr{\kamionkowski}.

\bigskip
\centerline{\bf 7. Baryonic Dark Matter (Machos)}
\smallskip
Probably the most exciting development in the dark matter story
is the detection of Machos by three separate groups
\rrr{\Nature,\EROS,\OGLE}
All three groups monitored millions of
stars \rrr{\Pac},
either in the LMC or in the galactic bulge, for signs
of gravitational microlensing, and all three groups have found it.
It has now become clear that these objects constitute
some new component of the Milky Way, but they do not constitute the
bulk of the dark matter.  Thus, the Macho search results
gives strong impetus to the search for particle dark matter.
However, the more than 60 detected microlensing events
are far in excess of predictions of standard Galactic models and
imply that the Galaxy is probably
quite different than was thought previously.
\bigskip

\line{\bf 7.1 Microlensing\hfil}
\smallskip

Microlensing has arrived as a powerful new tool for exploring
the structure of our Galaxy.
However, from the dark matter point of view, I'd like to note
that the current experiments may have the capability to
give a definitive answer to the question of whether the dark matter
in our Galaxy is baryonic.
The microlensing searches are probably sensitive to
any objects in the range $\sim 10^{-8}\msun < m < 10^3 \msun$,
just the range in which such objects are theoretically allowed to
exist.  Objects made purely of H and He with masses
less than $\sim 10^{-9} -10^{-7}\msun$ are expected to evaporate
due to the microwave background in less than a Hubble time, while
objects with masses greater than $\sim 10^3\msun$ would
have disrupted known globular clusters.

The idea of microlensing rests upon Einstein's observation that if a massive
object lies directly on the line-of-sight to a much more distance star,
the light from the star will be lensed and form a ring around the lens.
This ``Einstein ring" sets the scale for all the microlensing searches,
and in the lens plane, the radius of that ring
is given by
$$ r_E = 610 R_\odot [{m\over\msun} {L\over{\rm kpc}} x(1-x)]^{1/2},
$$
where $R_\odot$ and $\msun$ are the solar radius and mass, $m$
is the Macho mass, $L$ is the distance to the star being monitored,
and $x$ is the distance to the Macho divided by $L$.

In fact, it is extremely unlikely for a Macho to pass precisely
on the line-of-sight, but if there is a near miss, two
images of the star appear separated by a small angle.  For
masses in the stellar range and distances of galactic scale
this angle is too small to be
resolved, but the light from both images add and the star appears
to brighten.  The amount of brightening can be large, since it
is roughly inversely proportional to the minimum impact parameter $b/r_E$.
Since the Macho, Earth, and source star are all in relative motion,
the star appears
to brighten, reaches a peak brightness, and then fades back to its
usual magnitude.
The brightening as a function of time is called the ``lightcurve"
and is given by
$$
A(u) = {{u^2+2}\over{u \sqrt{u^2+4}}}, \qquad  u(t) =
[u_{min}^2 + (2 (t-t_0)/\that)^2]^{1/2},
$$
where $A$ is the magnification, $u=b/r_E$ is dimensionless impact
parameter, $t_0$ is the time of peak amplification, $\that = 2r_E/v_\perp$
is the duration of the microlensing event, $v_\perp$ is the transverse
speed of the Macho relative to the line-of-sight,
and $u_{min}$ is value of $u$ when $A=A_{max}$.
Thus the signature for a microlensing event is a time-symmetric
brightening of a star occurring as a Macho passes close to the line-of-sight.
When a microlensing event is detected, one fits the lightcurve and
extracts $A_{max}$, $t_0$, and $\that$.  The primary physical information
comes from $\that$, which contains the Macho velocity, and through
$r_E$ the Macho mass and distance.  Unfortunately, one cannot uniquely
find all three pieces of information from the measurement of $\that$.
However, statistically, one can use information about the halo density
and velocity distribution, along with the distribution of measured
event durations to gain information about the Macho masses.  Using a
standard model of the dark halo, Machos of jupiter mass ($10^{-3}\msun$)
typically last 3 days, while brown dwarf mass Machos ($0.1 \msun$)
cause events which last about a month \rrr{\pac,\Griest}.

In order to perform the experiment, a large number of stars must be followed,
since, assuming a halo made entirely of Machos, the probability of any Macho
crossing in front of a star is about $10^{-6}$.  Thus many millions of stars
must be monitored in order to see a handful of microlensing events.
In addition, if one wants to see microlensing from objects in the dark
halo, the monitored stars must be far enough away so that there is a lot
of halo material between us and the stars.  Therefore, the best stars to
monitor are those in the Large and Small Magellanic Clouds (LMC and SMC)
at distances of 50 kpc, and 60 kpc respectively, and also stars
in the galactic bulge, at 8 kpc.

There are several experimental groups that have undertaken the search
for microlensing in the LMC and galactic bulge
and have returned results.  The EROS collaboration,
has reported 3 events towards the LMC \rrr{\EROS}, the OGLE group has
reported about 15 events towards the bulge \rrr{\OGLE},
and the  DUO collaboration
has about a dozen preliminary events towards the bulge \rrr{\DUO}.
Our collaboration has seen about 5 events towards the
LMC \rrr{\Nature,\PRL,\LMCone},
and about 60 events towards the bulge \rrr{\Bulgeone,\Bennett,\Bulgetwo}.
We are also monitoring the SMC, but have yet to analyze
that data.  In what follows I will concentrate on MACHO collaboration data.

\bigskip
\line{\bf 7.2 The MACHO Collaboration Experiment \hfil}
\smallskip

The MACHO experiment is led by Charles Alcock and is a collaboration
of Physicists and Astronomers from Lawrence Livermore National Lab,
The UC Berkeley Center for Particle Astrophysics, Mount Stromlo and
Siding Spring Observatory, The University of Washington, Oxford, McMaster,
and UC San Diego.
We have essentially
full-time use of the $1.27$-meter telescope at
Mount Stromlo Observatory, Australia, for a period of about
8 years from July 1992.  In order to maximize throughput
a dichroic beamsplitter and filters provide simultaneous
images in two passbands, a `red' band (approx. 5900--7800 \AA)
and a `blue' band (approx. 4500--5900 \AA).
Two very large CCD cameras \rrr{\Stubbs}
are employed at the two foci;
each contains a $2 \times 2$ mosaic of $2048 \times 2048$ pixel
Loral CCD imagers, giving us a sky coverage of 0.5 square degrees.
Observations are obtained during all clear nights and part nights,
except for occasional gaps for telescope maintenance.
The default exposure times are 300 seconds for LMC images, 600
sec for the SMC
and 150 seconds for the bulge, so over 70 exposures are
taken per clear night.

Photometric measurements from these images are made with a
special-purpose
code known as SoDoPHOT \rrr{\bennettetal}, derived from DoPHOT
\rrr{\Schechter}.
For each star, the estimated magnitude and error are determined,
along with 6 other parameters (quality flags) measuring, for example,
the crowding, and the $\chi^2$ of the point-spread-function fit.
It takes about an hour on a Sparc-10 to process a field with 500,000 stars,
and so with the computer equipment available to us we manage to keep up.
The set of photometric data points
for each field are re-arranged into a time-series for each star,
combined with other relevant information including the seeing
and sky brightness, and then
passed to an automated
analysis to search for variable stars and microlensing
candidates \rrr{\Griesttime}.
The total amount of data collected to date is more than two Terabytes,
but the time-series database used for analysis is only about 100 Gbytes.

\bigskip
\line{\bf 7.3 Event Selection \hfil}
\smallskip

Most of the stars we monitor are constant within our photometric errors,
but about one half of one percent are variable.  The MACHO
database, as repository for the largest survey
ever undertaken in the time domain, is an extremely
valuable resource for studies of variable stars.  From our first year
LMC data alone we have already identified about 1500 Cepheid variables,
8000 RR Lyrae, 2200 eclipsing binaries, and 19000 long period variables.
Example lightcurves from each of these classes can be found in
reference \rrr{\Cook,\ourbinary}.
We also have many rare variables, and have given the first conclusive
evidence of 1st overtone pulsation in classical Cepheid's \rrr{\Ceph}.
We have also observed what may turn out to be entirely new types
of variable stars \rrr{\Cook}.

Given that the incidence of stellar variability, systematic error, and
other sources of stellar brightening is much higher than
the incidence of microlensing,
how can one hope to discriminate the signal from the background among the
tens of millions of stars we monitor nightly?
Fortunately, there are several very powerful microlensing
signatures which exist:
\item{1.}
High amplification.  Very high amplifications are possible, so we can
set our $A_{max}$ threshold high enough to avoid many types of systematic
error background.
\item{2.)}
Unique shape of lightcurve.  Only 5 parameters are needed to completely
specify the 2-color lightcurve.
\item{3.)}
Achromaticity.  Lensing magnification should be equal at all wavelengths,
unlike brightenings caused by most types of stellar variability.
\item{4.)}
Microlensing is rare.  The chance of two microlensing events occurring on
the same star is so small, that any star with more than one ``event"
can be rejected as a microlensing candidate.
\item{5.)}
Statistical tests:  The distribution of peak magnifications $A_{max}$ is
known a priori.  Microlensing should occur with equal likelihood on
every type and luminosity of star, unlike known types of stellar
variability.  New microlensing events should be discovered each year
at a constant rate.
\item{6.)}
Alert possibility.  Our alert system is now working and we can catch
microlensing before the peak and get many measurements of high accuracy.
Other spectral and achromaticity tests can also be performed in follow-up
mode.

Using these criteria, as well as others, we have found it possible to
pick out microlensing candidates from variable stars, etc.
For example, starting with about 9.5 million lightcurves from our first
year LMC database, we remove all but 3.
These are shown in Figure~7.

One of these events is clearly superior in signal/noise to the others,
and we have confidence in the microlensing label.
It has $A_{max}=7.2$, and $\that=35$ days.  The other two, while
passing all our cuts, and certainly consistent with microlensing,
are less certain to be actual microlensing.
We should note that our alert system has found a couple more high signal/noise
LMC microlensing events, which are not included here, since we have
performed efficiency calculations only on the first year data set.

Now, if we had found only these 3 events towards the LMC, we would
not be as confident as we are, that we have seen microlensing.
However, we have many more events towards the galactic bulge, and some of these
are of incredibly high signal/noise.
We cannot use the same selection criteria for the bulge as for the LMC
since our observing schedule towards the bulge is different, and the bulge
stellar population, distance, crowding, and extinction are different, but
using the same statistics, we can make a similar selection procedure.
We find about 43 candidates in our first year data (and since then
a few dozen more in our alert system).
Examples of lightcurves from the bulge are shown in Figure~8.
Some of these events are truly beautiful, with durations of many months
and magnifications of almost 20.
Coupled with the dozen events from the OGLE collaboration, I think
little doubt remains that microlensing has been seen.
\bigskip

\line{\bf 7.4 Detection Efficiency \hfil}
\smallskip

What do the microlensing events mean for the dark matter question?
In order to answer, we need to know the efficiency with which our
system can detect microlensing.  This is a non-trivial calculation.
In order to calculate our efficiencies,
we add simulated stars to real images, and then
artificially brighten them.  We run the photometry code on these simulated
images and find what the photometry code returns when a star brightens
under different atmospheric and crowding conditions.  These results are
incorporated into a large Monte Carlo in which simulated microlensing events
are added to our actual (non-microlensing) data
and fed into the same time-series analysis and selection procedure which
produced the 3 LMC microlensing candidates.
Thus we have explicitly taken into account inefficiencies caused by
bad weather and system down time, our analysis and selection procedure, as well
as blending of the underlying stars due to crowding, and systematic
errors in our photometry.
Since in order to calculate the expected number of events, we need to
integrate a theoretical microlensing rate over our measured
efficiency, we need efficiency $\epsilon$, as a function of $\that$.
The function $\epsilon$ can be found in reference \rrr{\PRL,\LMCone}.
Once $\epsilon$ is calculated, the
number of expected events is
$ N_{exp} = E \int {d\Gamma\over d\that} \epsilon(\that) d\that$,
where our total exposure $E_{LMC}=9.7 \ten{6}$ star-years, and
$d\Gamma/d\that$ is a differential microlensing rate which can be calculated
given a model of the dark halo \rrr{\Griest,\Explore}.
\bigskip

\line{\bf 7.5 Interpretation of LMC Events \hfil}
\smallskip

Using our sample of microlensing events, there are
two complementary analyses which can be performed.  First, we can set a
conservative limit on the Macho contribution to the dark halo.  Since
we know our efficiencies, and we have certainly not seen more than 3
microlensing events from halo objects,
any halo model which predicts more than 7.75 events can be ruled out
at the 95\% C.L.  This result will be independent of whether or not all
three candidate events are due to microlensing, and independent of whether
or not the lenses are in the dark halo.
Second, if we make the further assumption that all three events are due
to microlensing of halo objects, we can estimate the mass of the Machos
and their contribution to the mass of the dark halo.

In order to do either analysis we need a model of the dark
halo. We need to know the total mass of the halo, and we need the density
and velocity distribution
to calculate an expected microlensing rate.  The main constraints on
the halo come from the Milky Way rotation curve, which is not as well
determined as rotation curves in other galaxies.  Constraints from the orbits
of satellite galaxies also exist, but
there is considerable uncertainty in both the total halo mass and the
expected microlensing rate coming from uncertainty in the size and shape
of the Milky Way halo \rrr{\Gates,\Sackett,\Explore}.
Using a very simple, but commonly
used halo model \rrr{\Griest}, we can calculate the
number of expected events as
described above, and the results are shown in Figure~9.
If the Milky Way has a
standard halo consisting entirely of Machos of mass $0.001 \msun$
then we should have seen more than 20 events, with fewer events at larger
or smaller masses.  However, even if the halo dark matter consists
of Machos, it is very unlikely that
they all have the same mass.  Fortunately, it can be shown \rrr{\Griest}, that
if one rules out all halos made of unique Macho mass between masses
$m_1$ and $m_2$, then one has ruled out a halo consisting of ANY
distribution of masses as long as only masses between $m_1$ and $m_2$ are
included.
Thus we can make the powerful conclusion that a standard halo consisting
of any objects with masses between $8\ten{-5}\msun$ and $0.3\msun$
has been ruled out by our first year LMC data.

As mentioned above, there is no strong reason to believe that the Milky
Way halo is precisely as specified in the standard halo, and we would like
to test the robustness of the important results above by considering
a wider range of viable halo models.   To this end, we have investigated
a class of halo models due to Evans \rrr{\Evans}.  These
models have velocity distributions which are consistent with their density
profiles, and allow for halos which are non-spherical, and which have
rotation curves which gently rise or fall.  A description of the parameters
that specify these models, along with microlensing formulas
can be found in Alcock, \etal\ \rrr{\Explore}.
Basically we consider models which give rotation velocities within 15\%
of the IAU standard value of 220 km/sec, at the solar circle (8.5 kpc)
and twice the solar circle.  As pointed out by Evans
and Jijina \rrr{\Evansjijina},
the contribution of the stellar disk plays an important role in the predicted
microlensing rate.
This is because much (or even most) of the rotation
speed could be due to material in the disk, so we consider various size disks,
as well.

Using these models, we find
strong limits are found on heavy halo models, while only very
weak limits are found on light halo models.
This is because microlensing is sensitive not to the total mass in the halo,
but only to the mass in Machos.  So one can get a much more model independent
limit on the Macho content of the halo by limiting the {\it total mass
in Machos}, rather than the {\it Macho fraction} of the halo.
A more robust statement of our first year LMC microlensing results
is thus that objects in the $2\ten{-4} - 2\ten{2} \msun$ range can contribute
no more than $3 \ten{11}\msun$ to the dark halo, where we consider
the halo to extend out to 50 kpc.  The standard halo has $4.1 \ten{11}\msun$
out to this radius, and so is ruled out as before, but much smaller, all
Macho halos, would be allowed.  It should be clear that in order to
get good information on the Macho fraction of the halo, more
work is needed on the total mass of the halo.  This requires better
measurement of the Milky Way parameters and rotation curve.  Microlensing
measurements themselves may also be able to
help \rrr{\Sackett,\Gates,\Explore}.

The limits above are valid whether or not the three events shown in Figure~7
are due to microlensing of halo objects.  However, if we make the additional
assumption that they are, we can go beyond limits and estimate the Macho
contribution to the halo, and also the masses of the Machos.
The results obtained, especially on the lens masses, depend strongly on
the halo model used, so keep in mind that
it is not clear that all three events are microlensing, and it is certainly
not known that they are due to objects residing in the galactic halo.
Proceeding anyway, we can construct a likelihood function as the product
of the Poisson probability of finding 3 events when expecting
$N_{exp}$ and the probabilities of drawing the observed $\that$'s
from the calculated model duration distribution \rrr{\Explore,\PRL,\LMCone}.
The resulting likelihood contours can be found in references \rrr{\PRL}
and \rrr{\LMCone}.
We find that for a standard
halo, a macho fraction of $\sim 20$\% is most likely, with Macho
masses in the $0.01 - 0.1\msun$ range likely.  Note that the errors
in these estimates are very large due to the small number statistics,
and that there is an enormous additional uncertainty due to the halo
model.  However, once again, the maximum likelihood estimate of the total
mass in Machos is quite model independent and is about $8 \ten{10}\msun$.
Since the mass in known stars, gas, etc. is only about $6 \ten{10}\msun$,
we see this would be a major new component of the Milky Way if
it is confirmed to exist.

\bigskip
\line{\bf 7.6 Interpretation of Bulge Events \hfil}
\medskip

The large number of events we (and the OGLE group \rrr{\OGLE}) have found
towards the galactic center came as a great surprise to everyone.
The line-of-sight toward the bulge goes though the stellar disk,
so bulge microlensing is sensitive to halo dark matter, disk stars, and any
disk dark matter which might be present.
The early predictions \rrr{\Griestbulge,\Pacbulge,\Kiraga}
included all these sources,
but still predicted many fewer events than have now been observed.
It seems the microlensing experiments have
discovered a new component of the Milky Way.
A standard way of quoting the microlensing probability is the optical
depth $\tau$, which is the probability that any given source star is
lensed by a magnification of 1.34 or greater.  Optical depth has larger
statistical errors than the event rate, but has the great advantage
of being independent of the masses of the lenses.
Early predictions of bulge microlensing were in the $10^{-6}$
range \rrr{\Griestbulge,\Pacbulge,\Kiraga},
while using the sample of events above we find $\tau_{est} =
3.9 ^{+1.8}_{-1.2} \ten{-6}$ \rrr{\Bulgetwo}.
We have not finished the complete
efficiency calculation for our bulge events, so this estimate uses
a sub-sample of 15 giant star events, for which our preliminary
efficiencies should be acceptable \rrr{\Bulgetwo}.

Several models have now
been proposed to explain the high microlensing rate.
They include \rrr{\Gould, \Zhou, \Pacc}
\item{1)}
A ``heavy" disk.  Perhaps the disk of the Milky Way is substantially
more massive than normally considered.

\item{2)}  A ``bar" at small inclination.
Perhaps the Milky way is not a grand design spiral
as usually assumed, but is a barred spiral, with a very large bar, previously
overlooked since it points nearly toward us.

\item{3)}  A highly flattened, or disk-like halo.

\item{4)} Some combination of the above, and/or extra material in the bulge.

The suggestion of a Galactic bar has been around for a few years,
and seems to be corroborated by other data \rrr{\Spergelbar},
though it is still
not clear whether this alone is sufficient to explain the microlensing data.
Extensive work is being undertaken in trying to resolve these questions.
One method is to map out the bulge area with microlensing.  A bar-like
structure will give a different pattern of microlensing than a disk-like
structure.  Use of a satellite, or the fine-structure of the microlensing
lightcurve has also been suggested \rrr{\gouldparallax,\gouldsatellite}.
\bigskip

\line{\bf 7.7 Advantages of Having Many Events \hfil}
\smallskip

There are two main advantages of having several times
more events than we originally thought we would have.
First, we can do statistical tests on the data.
For example, simple geometry predicts a specific distribution of maximum
amplifications.  Basically, every lens/line-of-sight impact parameter
should be equally likely, so the distribution of $u_{min}$'s should
be uniform (taking into account that our efficiency for detecting high
magnification (low $u_{min}$) events is larger).
We have performed a Kolmogorov-Smirnov (K-S) test on the bulge events
and find good consistency with the microlensing hypothesis.  Thus
the microlensing interpretation of these events is greatly strengthened.

The second advantage of having many events, is that rare events can be found.
For example events of high magnification or long duration
should occur occasionally.
For some types of rare events additional important information concerning
the Macho mass/velocity/distance can be extracted.
For example, in reference \rrr{\parallax}
we show an event which lasted about 1/2 year, during which time
the Earth had a chance to travel part way around the Sun.  This gave
our telescope two different perspectives on the lens, resulting in a
parallax event.  Thus the lightcurve does not fit the naive amplification
formula presented earlier.
Including the Earth's motion, we find
a good fit, and discover that the Macho was moving with
a projected transverse velocity of $76 \pm6$ km/sec.  The Macho mass
is determined by a combination of this velocity, the event duration,
and the distance to the Macho, so for such parallax events there is a
one-to-one relationship between the Macho mass and distance.  In this
case the Macho could be either a brown dwarf star in the galactic bulge,
an M-dwarf star at a distance of 2 to 6 kpc, or a more massive star
quite nearby.

Another rare type of microlensing event is shown in
reference \rrr{\ourbinary,\Bennett}.
This lightcurve is characteristic of lightcurves formed by binary lenses.
This particular event was first seen by the OGLE group \rrr{\OGLE},
and detailed analysis will again give information as to the
lens masses, distances
and velocities.  An exciting aspect of such a binary Macho detection,
is the possibility of detecting planets around Machos.  Given that some
of the lenses we observe are in fact low mass stars, it is possible
to observe caustic crossing such as mentioned above, for
planets even down to Earth mass \rrr{\Mao,\Loeb}.
Thus microlensing may well be the best
way to discover and get statistics on extra-solar planets.
\bigskip

\line{\bf 7.8 Macho Conclusion \hfil}
\smallskip

The microlensing experiments have given robust and strong limits on the
baryonic content of the halo.  Much more data from the LMC and SMC will
be available soon, so we expect the statistics to improve in the near
future.  The LMC events, if interpreted as due to halo microlensing,
allow a measurement of the baryonic contribution to the halo, which is
around 20\% for a standard halo.  In this case, the most likely
Macho contribution to the Milky Way halo mass is about $8\ten{10}\msun$,
which is roughly the same as the disk contribution to the Milky
Way mass.  However, the whole story has been made
more complicated (and exciting)
by the much larger than expected number of bulge microlensing events.
These events imply a new component of the Galaxy, and until the nature
of this new component is known, unambiguous conclusions concerning the
LMC events will not be possible.  For example, if the Milky Way disk
is much larger than usually considered, a much smaller total halo mass
will be required, and so even an all Macho halo might be allowed.
Alternatively, the new Galactic component which is giving rise to the
bulge events, may also be giving rise to the LMC events, and the Macho
content of the halo could be zero.
Fortunately, much more data is forthcoming, and many new ideas have been
proposed.  Microlensing is fast becoming a new probe of Galactic structure,
and beside the original potential to discover or limit dark matter,
may well produce discoveries such as extra-solar planetary systems.
\bigskip

\centerline{\bf 8. Conclusions}
\smallskip

The dark matter situation has changed dramatically in the past few years.
Not long ago, people agreed that the dark matter existed, but had little
hope of knowing what it actually consisted of.  Now strong detection
efforts are underway for many of the best candidates.  For Machos,
first results are already in, and it seems quite probable that the bulk
of the dark matter does not consist of Machos in the Earth to
brown dwarf mass range.  There is still a ``baryonic dark matter"
window for exotic objects in the solar to 1000 solar mass range.
Turning to Wimps, we found that these are excellent dark matter candidates
for a variety of reasons, and that three methods of detection
are being vigorously pursued:  high energy accelerators, direct detection,
and via high energy neutrinos from the Sun.
Axions also are fine dark matter candidates, and the new microwave
cavity experiments will for the first time probe some of the best
axion parameter space.  However, no experiments capable of definitively
ruling out either axion or Wimp candidates are underway, so
there is the chance that either could be the dark matter without
us discovering it.

No one has yet found a method to directly
detect a light neutrino component of the dark matter, though interest
in these as candidates for the ``hot dark matter" component in
a mixed hot plus cold dark matter galaxy formation scenario is very high.
For neutrinos, the most promising method is to measure the masses
via neutrino oscillation experiments, and then calculate the relic
density using the big bang theory.  Indirect and preliminary
evidence for such neutrino oscillations already exists, so experiments
capable of actually determining neutrino masses should be watched
with great interest by all astrophysicists.

In conclusion, this is a very active field, and remarkably, there
is a reasonable chance of discovering the nature of the dark matter
within the next few years.

We thank Andrew Gould, George Fuller, Joel Primack, and members
of the MACHO collaboration for valuable help.
We acknowledge support from a DoE OJI Award,
the NSF Center for Particle
Astrophysics (AST-8809616), the Alfred P. Sloan foundation, and a Cotrell
Scholars Award.

\bigskip

%\par \penalty-400 \vskip\chapterskip
   %\spacecheck\referenceminspace
   \immediate\closeout\referencewrite
   \referenceopenfalse
   \line{\bf\hfil References\hfil}%\vskip\headskip
   \input referenc.texauxil
   
\vfill
\eject
\centerline{\bf Figure Captions}
\medskip

{\bf Figure 1. }
     Parameter space in the minimal supersymmetric model.
     Only two of the five dimensions
     ($\mu$ and $M_2$) are displayed.  Panel (a) shows the starting
     grid of parameters choices, and panel (b)
     shows the models left after eliminating models which violate
     any of several accelerator constraints (from \rrr{\Jungman}).

{\bf Figure 2.}
	Scatter plot of
        relic neutralino density vs. neutralino mass for the set
        of supersymmetric models discussed in the text.
        Laboratory constraints from LEP measurements and
        Br($b\rightarrow s\gamma$) are enforced.   Models between
        the lines drawn at $\Omh=0.025$ and $\Omh=1$ are compatible with
        neutralino dark matter (from \rrr{\Jungman}).

{\bf Figure 3.}
  Predicted rate in a $^{73}$Ge cryogenic detector vs neutralino
     mass for the allowed dark-matter models above  (from \rrr{\Jungman}).

{\bf Figure 4.}
     Measured gamma-ray background in an underground high-purity
     germanium ionization detector (data acquired by the
     UCB/UCSB/LBL experiment at Oroville \rrr{\angela,\caldwell}.
     Various gamma-ray lines are identified, as is the end point
     of the broad tritium spectrum.  The rapid rise at low $Q$
     is the electronic noise (from \rrr{\Jungman}).

{\bf Figure 5.}
     Theoretical differential event rate for WIMPS of various
     masses hitting a germanium target.  WIMP masses are labeled
     in GeV. An arbitrary cross section of  $\sigma_0=4 \times
     10^{-36} {\rm cm}^2$ was chosen with standard values for
     the other parameters.
     Note the rate axis scale is 100 times {\it smaller} than in
     Fig.~4,  % {\bf HARDWIRED},
     and the cross section chosen is very high for neutralinos
	(from \rrr{\Jungman}).

{\bf Figure 6.}
     Indirect-detection rate vs neutralino mass.  The sum of the rates
     for upward muons from both the Sun and Earth is shown.  Currently
     planned experiments will be
     sensitive in the $10^{-2}\inunit$ to $10^{-4}\inunit$ range
     (from \rrr{\Jungman}).

{\bf Figure 7.}
The three observed stellar lightcurves
that we interpret as gravitational microlensing events are each shown in
relative flux units (red and blue) vs time in days.
The solid lines are fits to the theoretical microlensing shape
(from \rrr{\PRL}).

{\bf Figure 8.}
Example lightcurves from first year bulge data.  Four of the 43 microlensing
events are shown (from \rrr{\Cook,\ourbinary}).

{\bf Figure 9.}
The lower panel shows the number of expected events predicted
from the standard model halo with a delta function mass distribution.
Given three observed events, points above the line drawn at $N_{exp}=7.7$
are excluded at the 95\% CL.   The upper panel shows the 95\% CL limit on the
halo mass in MACHOs within 50 kpc of the galactic center
for the model.  Points above the curve are excluded at 95\%
CL while the line at $4.1 \ten{11} \msun$ shows the total mass in this
model within 50 kpc  (from \rrr{\PRL}).

\vfill
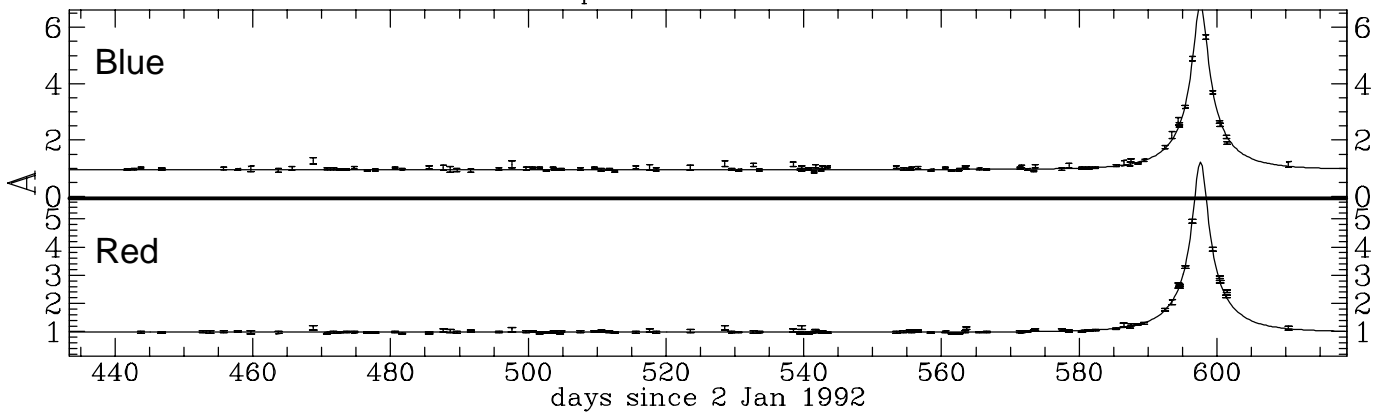
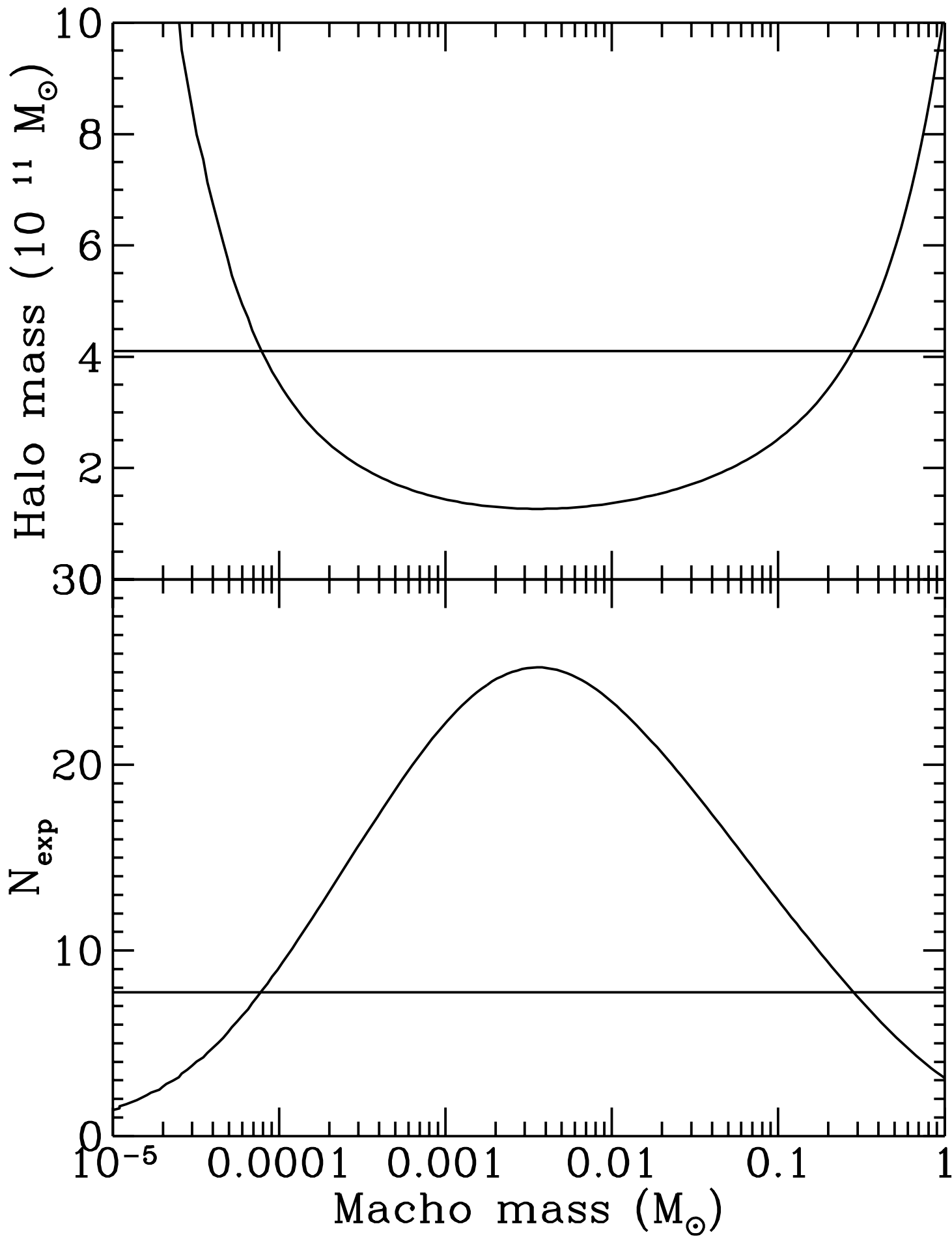
\end